\documentclass[10pt]{aastex}
\usepackage{emulateapj5,apjfonts}
\usepackage{epsf}
\usepackage{graphicx}
\slugcomment{{\sc Submitted to ApJ:} January 23, 2004}
\newcommand{\begit}{\begin{itemize}}
\newcommand{\enit}{\end{itemize}}
\newcommand{\begen}{\begin{enumerate}}
\newcommand{\enen}{\end{enumerate}}
\newcommand{\beq}{\begin{equation}} 
\newcommand{\eeq}{\end{equation}} 
\newcommand{\beqa}{\begin{eqnarray}} 
\newcommand{\eeqa}{\end{eqnarray}} 
\newcommand{\p}{\partial}    
 
\newcommand{\bof}{B_{0_{15}}}

\newcommand{\rnuten}{R_{\nu_{10}}}

\newcommand{\mdotth}{\dot{M}_{-3}}
\newcommand{\mns}{M_{1.4}}

\begin{document} 

\title{Magnetar Spindown, Hyper-Energetic Supernovae, \& Gamma Ray Bursts}

\author{Todd A. Thompson\altaffilmark{1,2}, Philip Chang\altaffilmark{3}, \& Eliot Quataert\altaffilmark{4}}
\altaffiltext{1}{Hubble Fellow}
\altaffiltext{2}{Astronomy Department 
and Theoretical Astrophysics Center, 601 Campbell Hall, 
University of California, Berkeley, CA 94720; 
thomp@astro.berkeley.edu}
\altaffiltext{3}{Department of Physics, Broida Hall, University of California, 
Santa Barbara, CA 93106; pchang@physics.ucsb.edu}
\altaffiltext{4}{Astronomy Department 
and Theoretical Astrophysics Center, 601 Campbell Hall, 
University of California, Berkeley, CA 94720; 
eliot@astro.berkeley.edu}
\begin{abstract}

The Kelvin-Helmholtz cooling epoch, lasting tens of seconds after the 
birth of a neutron star in a successful core-collapse supernova, is accompanied by
a neutrino-driven wind.  For magnetar-strength ($\sim10^{15}$ G) large scale
surface magnetic fields, this outflow is magnetically-dominated during the entire cooling epoch.
Because the strong magnetic field forces the
wind to co-rotate with the protoneutron star,
this outflow can significantly effect the neutron star's early angular momentum evolution,
as in analogous models of stellar winds (e.g.~Weber \& Davis 1967). 
If the rotational energy is large in comparison with the supernova energy
and the spindown timescale is short with respect to the time required for the
supernova shockwave to traverse the stellar progenitor, the energy extracted may
modify the supernova shock dynamics significantly.  This effect is capable
of producing hyper-energetic supernovae and, in some cases, provides
conditions favorable for gamma ray bursts.  We estimate spindown timescales
for magnetized, rotating protoneutron stars and construct steady-state models of
neutrino-magnetocentrifugally driven winds.   We find that if magnetars are born rapidly
rotating, with initial spin periods ($P$) of $\sim1$ millisecond, that of order $10^{51}-10^{52}$ erg
of rotational energy can be extracted in $\sim10$ seconds. If magnetars are born slowly
rotating ($P\gtrsim10$ ms) they can spin down to periods of $\sim1$ second on the 
Kelvin-Helmholtz timescale.  
\end{abstract}

\keywords{stars: magnetic fields --- stars: winds, outflows --- stars: neutron --- supernovae: general --- gamma rays: bursts}

\section{Introduction}

A successful supernova leaves behind a hot deleptonizing protoneutron star (PNS).
This newly born, contracting, compact object radiates its gravitational binding
energy in neutrinos, which ablate matter from its surface via energy deposition.  
The primary heating mechanisms are the charged-current processes of electron and
anti-electron neutrino absorption on free nucleons: $\nu_e n\rightarrow p e^-$ and
$\bar{\nu}_e p\rightarrow n e^+$.  The neutrino energy (typically $\sim10$ MeV)
is deposited in the final-state leptons, which are hydrodynamically coupled to
the nucleons.  In this way, a thermal wind composed of nucleons and leptons is created.
At the densities and temperatures surrounding the PNS at birth, photons are completely trapped
and are advected with the wind (e.g.~Duncan et al.~1986; Qian \& Woosley 1996).

This neutrino-driven outflow emerges into the supernova post-shock environment --
the supernova shock itself moving away from the protoneutron star at a velocity of 10-30,000 km s$^{-1}$
(e.g.~Burrows, Hayes, \& Fryxell 1995).
The duration of the PNS wind is set by the cooling
or Kelvin-Helmholtz timescale ($\tau_{\rm KH}\sim GM^2/L_\nu R$, where $L_\nu$ is the total neutrino
luminosity) for radiating away heat and lepton number and is typically of order tens of seconds
(Burrows \& Lattimer 1986; Pons et al.~1999).  The kinetic luminosity of the wind is typically
less than 10$^{48}$ erg s$^{-1}$ and decreases sharply as the neutrino luminosity decreases
(e.g.~Qian \& Woosley 1996)
and the protoneutron star cools, and so the addition to the asymptotic supernova
energetics is small on the scale of the canonical supernova energy, $10^{51}$ erg.  
Furthermore, the total mass ejected during the cooling epoch is 
$\lesssim10^{-3}$ M$_\odot$, depending upon how the start of the wind phase is defined -- minor in
comparison with the total mass ejected in a typical core-collapse supernova.
Despite this status as a mere perturbation to supernovae in both mass and energy,
PNS winds have been the focus of considerable recent work. 
In particular, because of their intrinsic neutron-richness and association with supernovae,
research on PNS winds has focused on the possibility that they
might be the as yet unidentified astrophysical site for production of the $r$-process nuclides (Woosley et al. 1994; 
Takahashi et al.~1994; Qian \& Woosley 1996; Cardall \& Fuller 1997; Sumiyoshi et al. 2000; 
Otsuki et al. 2000; Wanajo et al. 2001; Thompson et al. 2001).   

Recently, Thompson (2003a,b) noted that in the absence of rotation magnetar-like 
(e.g.~Kouveliatou et al.~1999; Duncan \& Thompson 1992; Thompson \& Duncan 1993)  
surface magnetic field strengths ($B_0\sim10^{15}$ G) can dominate the thermal pressure 
and kinetic energy density of the wind in this very early
phase of neutron star evolution.  
Here we extend this work by considering in a simplified model the combined 
action of neutrino heating, rotation, and strong magnetic fields.  

Magnetic dipole radiation (e.g.~Pacini 1967, 1968; Gunn \& Ostriker 1969) 
and, to a lesser extent, gravitational wave radiation (e.g.~Ostriker \& Gunn 1969) are often
considered as dominant spindown mechanisms, controlling the rate of rotational energy extraction
and the angular momentum evolution of young neutron stars.  Here we consider a mechanism
well-known to the solar physics community: magnetic braking by the combined action of
a persistent outflow and a strong magnetic field (Weber \& Davis 1967; Mestel 1968; Belcher \& Macgregor 1976;
Hartmann \& Macgregor 1982; Mestel \& Spruit 1987).  For a wide range of initial rotation periods,
we find that protoneutron stars may be spun down by the presence of this neutrino-magnetocentrifugally
driven outflow in the first tens of seconds of the PNS's life.  
If these highly magnetic objects are born rapidly rotating, $\sim10^{52}$ erg may be
extracted on the cooling timescale, giving an energetic boost to the just-preceding supernova and providing
favorable conditions for gamma ray bursts. 

Although our focus is on magnetars born with millisecond periods, we also assess the evolution
of highly magnetic objects born slowly rotating, with spin
periods similar to those of radio pulsars (tens to hundreds of milliseconds).
Our conclusions for these objects are sensitively dependent upon
magnetic field geometry.  However, for monopole-like fields we come to the interesting conclusions that
these objects are spun down considerably on the cooling
timescale; a PNS with $\sim20$\,ms initial period is transformed 
into a $\sim1$ second rotator in $\tau_{\rm KH}$.

\subsection{This Paper}

In \S\ref{section:motive} we review the basics of stellar spindown via magnetocentrifugal outflows and
provide some motivation for pursuing this effect in PNS winds.
Section \ref{section:eta} discusses some aspects and ambiguities of the expected magnetic field topology.
In \S\ref{section:analytic} we provide simple analytic scalings for the Alfv\'{e}n point
and spindown timescales over a wide range of initial PNS spin periods.  Section \ref{section:nmcwind}
describes our models of neutrino-magnetocentrifugally-driven outflows 
and presents estimates of rotational energy extraction.  
In \S\ref{section:implications} we discuss applications to hypernovae and gamma ray bursts,
making contact with Usov (1992) and Thompson (1994).  We further
summarize our findings, review the potential implications for supernova
remnants and the angular momentum evolution of young neutron stars,
and highlight open questions in need of further investigation.
Throughout this paper we refer to models
of non-rotating, non-magnetic PNS winds (e.g. the models of 
Duncan et al.~1986; Takahashi et al.~1994; Qian \& Woosley 1996; 
Sumiyoshi et al.~2000; Otsuki et al.~2000; Wanajo et al.~2001;
Thompson et al.~2001) with the acronym `NRNM'.  

\section{Spindown}
\label{section:motive}

Mass loss from the surface of any rotating star carries away angular momentum.  
If the star has a strong magnetic field, the matter lost in the stellar wind is forced
into near corotation with the stellar surface out to $\sim R_{\rm A}$, 
the Alfv\'{e}n point, where the magnetic energy density equals the kinetic energy 
density of the outflow.  This effect provides for efficient angular momentum transport
from the rotating star to the outflow (Schatzman 1962). Angular momentum conservation implies that
\beq
\frac{d}{dt}(I\Omega)=-\dot{M}{\cal L}
\label{nonrellcons}
\eeq
where $I$ is the moment of inertia, $\Omega$ is the angular velocity at the stellar surface, $\dot{M}$
is the wind mass loss rate, and ${\cal L}$ is the specific angular momentum carried by the wind.
In the classic model for solar spindown constructed by Weber \& Davis (1967), the problem is treated
in one spatial dimension at the equator with ${\bold B}=B_r {\bf \hat{e}}_r+B_\phi{\bf \hat{e}}_\phi$
and ${\bf v}=v_r{\bf \hat{e}}_r+v_\phi{\bf \hat{e}}_\phi$.  At the stellar surface $B_r\gg B_\phi$,
and the field is monopolar, $B_r\propto r^{-2}$.\footnote{This field configuration is often referred to as `split-monopole'.}
Consideration of the azimuthal momentum 
equation together with Faraday's law yields 
\beq
{\cal L}=rv_\phi-\left(\frac{rB_rB_\phi}{4\pi \rho v_r}\right)={\rm constant}=R_{\rm A}^2\Omega,
\label{angmom}
\eeq
where $\rho$ is the mass density.  Although the magnetic field lines bend in the $-\phi$
direction between the stellar surface and $R_{\rm A}$, such that $B_r(R_{\rm A})\sim B_\phi(R_{\rm A})$,
eq.~(\ref{angmom}) states that the wind has angular momentum as if strict co-rotation
is enforced between the magnetic field footpoints and $R_{\rm A}$.

Employing eq.~(\ref{angmom}) in eq.~(\ref{nonrellcons}) and taking $I=(2/5)MR_\nu^2$, 
where $M$ is the PNS mass and $R_\nu$ is the PNS radius\footnote{The subscript $\nu$ refers to
the fact that in the PNS case we take the `surface' to correspond to the radius
of neutrino decoupling, the neutrino-sphere.}, one finds that
\beq
\Omega_f=\Omega_i\left(M_f/M_i\right)^{\frac{5}{2}(R_{\rm A}/R_\nu)^2}.
\label{fracomegaloss}
\eeq
Here we have assumed that $R_\nu$ and $R_{\rm A}$ are independent of time.
The subscripts $f$ and $i$ denote `final' and `initial', respectively.
In the absence of magnetic fields, $R_{\rm A}=R_\nu$ and for a total mass loss of 
even $10^{-2}$ M$_\odot$, $\Omega$ changes by just 1\% for a solar mass neutron star.
For strong magnetic fields, however, we expect a large $R_{\rm A}/R_\nu$, and, correspondingly,
large $\Delta\Omega$.   If strongly magnetized PNSs are born rapidly rotating 
(Duncan \& Thompson 1992; Thompson \& Duncan 1993) then a reservoir of rotational energy,
\beq
E_{\rm Rot}=\frac{1}{2}I\Omega^2\sim\frac{1}{5}MR_\nu^2\Omega^2\simeq2.2\times10^{52}\,\,{\rm erg}\,\,\,
\mns\,\rnuten^2\,P_1^{-2},
\label{energy}
\eeq
large in comparison with $10^{51}$ erg, may be extracted on the spindown timescale.  
In eq.~(\ref{energy}), $M_{1.4}=M/1.4$ M$_\odot$, $R_{\nu_{10}}=R_\nu/10\,\,{\rm \,km}$, 
and $P_1$ is the spin period in units of 1 millisecond (ms).  The spindown timescale is
defined here as the $e$-folding time for $\Omega$ (combining eqs.~\ref{nonrellcons} and \ref{angmom});
\beq
\tau_{\rm J}=\frac{\Omega}{\dot{\Omega}}=\frac{2}{5}\frac{M}{\dot{M}}\left(\frac{R_\nu}{R_{\rm A}}\right)^2.
\label{spintime}
\eeq
If $\tau_{\rm J}$ is short compared with the Kelvin-Helmholtz cooling timescale for the PNS, then
the angular momentum of the PNS is significantly effected during the wind epoch.  Also,
because $E_{\rm Rot}$ is potentially much larger than $10^{51}$ erg, 
if $\tau_{\rm J}$ is short compared with the time for the just-preceding supernova shockwave
to traverse the progenitor,  then we expect  interesting observational consequences and modifications to
the supernova explosion, resulting nucleosynthesis,  and remnant dynamics.  The last of these,
that the supernova remnant may have larger inferred asymptotic kinetic energy is not unique to
this spindown mechanism.  Simple vacuum dipole spindown also energizes the supernova
remnant on a {\it relatively} short timescale for highly magnetic, rapidly rotating neutron stars.
The primary differences here is that if a large fraction of the rotational energy can be
tapped quickly, the nucleosynthesis of the actual explosion can be modified. This difference is
of particular importance because, as 
we show in \S\ref{section:analytic}, the spindown timescale given in eq.~(\ref{spintime})
can be significantly shorter than that inferred from vacuum dipole spindown.

\section{Field Topology}
\label{section:eta}

To calculate the spindown of PNSs, we
must estimate $R_{\rm A}$.  To do so, we appeal to its definition --
the point at which the magnetic energy density ($B^2/8\pi$)
equals the radial wind kinetic energy density ($\rho v_r^2/2$).
For the purposes of simplicity and generality, we take
\beq
B_r=B_0(R_\nu/r)^{\eta}.
\label{generalb}
\eeq
The Weber-Davis model posits a monopole field structure with $\eta=2$.
With this dependence of $B$ on $r$, we may estimate
$R_{\rm A}$ and $\tau_{\rm J}$.
We may rightly ask, however, whether or not these estimates are appropriate considering
the fact  that the currents in the star should necessarily produce -- to lowest order -- a dipole field
with $\eta=3$.  This question is germane because the predicted $R_{\rm A}$ 
is much smaller if one assumes a dipole field and, consequently, the inferred 
$\tau_{\rm J}$ is much longer.  The ambiguity is due to the intrinsically multi-dimensional
nature of magnetically-dominated outflows.  Analytical (e.g.~Mestel 1968; Mestel \& Spruit 1987) and numerical
models (e.g.~Pneumann \& Kopp 1971; Steinolfson et al.~1982; Usmanov et al.~2000; Lionello et al.~2002;
Ud-Doula \& Owocki 2002) show that the interaction between the magnetic field and the flow
is complex.  The steady-state structure, for modestly magnetically-dominated flows, is the classic helmet-streamer
configuration (see Fig.~1 of Mestel \& Spruit 1987 for a depiction) with magnetic field lines emerging
from low latitudes forming a `dead zone', a region of closed magnetic loops, and the wind 
at high latitudes opening the field to infinity.  The resulting field is a mix of  `monopole' and `dipole'.

Because the kinetic energy density of the outflow dominates the magnetic field for $r>R_{\rm A}$,
and vice-versa for $r<R_{\rm A}$, one might naively expect that inside $R_{\rm A}$ the field is
roughly dipolar and outside $R_{\rm A}$ it is monopolar.
This picture is the most conservative from the point of view of spindown and describes the 
basics of the model of Mestel (1968).  Rapid rotation complicates the issue.  Mestel \& Spruit (1987) show
that the radial extent of the dipolar region, the dead zone ($R_{\rm D}$), can be smaller than $R_{\rm A}$
(also implied by Ud-Doula 2002).
Assuming isothermal conditions, the radial extent of the dead zone can be calculated 
from the equations of magnetohydrostatic equilibrium;
\beqa
\left(\frac{R_\nu}{R_{\rm D}}\right)^6&=&\frac{\rho_{\nu} \,c_{T_\nu}^2}{(B_0^2/8\pi)}
\exp\left[-\frac{GM}{R_\nu c_{T_\nu}^2}\left(1-\frac{R_\nu}{R_{\rm D}}\right)\right] \nonumber \\
&\times&\hspace{1.28cm}\exp\left[
\frac{R_\nu^2\Omega^2}{2c_{T_\nu}^2}\left(\frac{R_{\rm D}^2}{R_\nu^2}-\frac{R_\nu}{R_{\rm D}}\right)\right]
\label{rd}
\eeqa
where $c_{T_\nu}$ and $\rho_\nu$ denote the isothermal sound speed and mass density, respectively,
in the dead zone, at $R_\nu$
Equation (\ref{rd}) is  a consequence of balancing magnetic tension with thermal pressure
in an isothermal magnetohydrostatic atmosphere.  Taking fiducial PNS parameters as in \S\ref{section:analytic}
($R_\nu=10$ km, $P=1$ ms, $B_0=10^{15}$ G, and $M=1.4$ M$_\odot$), $R_{\rm D}$
is typically $<20$ km, depending upon the choice for $\rho_\nu$ and $c_{T_\nu}$, 
much less than $R_{\rm A}$ (see eq.~\ref{eta2spin}).
In the range of radii $R_{\rm D}<r<R_{\rm A}$, the field lines are opened into a monopole configuration and
$R_{\rm A}$ is estimated with $\eta=2$ for the field structure.  This effect yields
spindown timescales  that are shorter than those inferred from the more pessimistic case
in which a dipole is assumed from the stellar surface all the way out to $R_{\rm A}$.
Therefore, eq.~(\ref{rd})  implies that for rapid rotation 
(where `rapid' is defined by the ratio $R^2_\nu\Omega^2/c^2_{T_\nu}$)
the net spindown might be better approximated by employing a pure monopole model,
whereas for slow rotation a dipole field is probably more appropriate.

In addition to this centrifugal effect, there are several other reasons to suspect that in estimating $R_{\rm A}$
we should take $\eta<3$ in eq.~(\ref{generalb}).  First, any net twist to the magnetic field lines yields $\eta<3$
and spindown is enhanced (Thompson et al.~2001). Second, the braking indices inferred from 
observations of pulsars, a system in which  the pure dipole limit should strictly obtain,
are inconsistent with $\eta=3$.  Third, the dead zone on the surface of a PNS may be periodically opened
by neutrino heating (Thompson 2003), footpoint motion and shear (Thompson \& Murray 2001), or MHD instabilities.   
In short, $\eta$ must be less than 3, but it is greater than 2.  
Both are limiting cases.
The complications of rapid rotation (eq.~\ref{rd}),
relativity (we encounter relativistic outflows in \S\ref{section:analytic}), and neutrino heating make this
ambiguity resolvable only with multi-dimensional numerical simulations of 
thermal magnetocentrifugal winds.

In what follows, we present estimates for arbitrary $\eta$ and scalings for $\eta=2$ and $\eta=3$.
As we have just argued, our results for the monopole case likely over-estimate the efficacy of spindown 
(Poe, Friend, \& Cassinelli 1989;  Keppens \& Goedbloed 1999; Van der Holst et al.~2002; {\it but}, see Pizzo et al.~1983).
However, the results from the pure dipole ($\eta=3$) limit must under-estimate  spindown (Mestel \& Spruit 1987).  

\section{Estimates}
\label{section:analytic}

At $R_{\rm A}$, $B^2/8\pi\sim\rho v_r^2/2$.  With eq.~(\ref{generalb}), and using $\rho=\dot{M}/4\pi r^2 v_r$,
\beq
R_A^{2\eta-2}=B_0^{2}\,R_\nu^{2\eta}\,\dot{M}^{-1}v_{\rm A}^{-1},
\label{slowspin}
\eeq
where $v_{\rm A}=v_r(R_{\rm A})$ is the Alfv\'{e}n speed.
Let $v_\nu$ denote the
asymptotic velocity attained by the matter in a purely NRNM outflow (that is, in a wind with $\Omega=0$ and $B_0=0$).
Because the flow in NRNM wind models is driven by extremely inefficient neutrino heating, $v_\nu$
is typically much less than the PNS escape speed (see e.g.~Thompson et al.~2001).  For high neutrino luminosity $v_\nu\lesssim3\times10^9$ cm s$^{-1}$.
If the matter is forced to co-rotate to $\sim R_{\rm A}$, then $v_\phi(R_{\rm A})\sim R_{\rm A}\Omega$. 
If  $R_{\rm A}\Omega\gtrsim v_\nu$\footnote{
This is equivalent to assuming the {\it fast magnetic rotator} (FMR) limit of Belcher \& Macgregor (1976).}  
we expect that $v_r(R_{\rm A})\sim v_\phi(R_{\rm A})$.  These arguments  imply 
that  to reasonable approximation $v_{\rm A}\sim R_{\rm A}\Omega$.  Therefore, from eq.~(\ref{slowspin}), 
\beq
R_A^{2\eta-1}=B_0^{2}\,R_\nu^{2\eta}\,\dot{M}^{-1}\Omega^{-1}.
\label{slowspin2}
\eeq
Scaling for a rapidly rotating magnetar ($P=1$ ms),
\beqa
R_A (\eta=2)&\simeq&43\,\,{\rm km}\,\,\,
\bof^{2/3}\,\rnuten^{4/3}\mdotth^{-1/3}\,P_{1}^{1/3}
\label{eta2spin} \\
R_A(\eta=3)&\simeq&24\,\,{\rm km}\,\,\,
\bof^{2/5}\,\rnuten^{6/5}\,\mdotth^{-1/5}P_{1}^{1/5}
\label{eta3spin}
\eeqa
Here and throughout this paper, $B_{0_n}=B_0/10^n$ G, $\dot{M}_{-m}=\dot{M}/10^{-m}{\rm\,\,M_\odot\,\,s^{-1}}$, 
and $P_l=P/l{\rm \,\,\, ms}$. 
Note that these scalings apply only when the condition $R_{\rm A}\Omega\gtrsim v_\nu$
is satisfied.\footnote{In the more general case ($v_\nu\sim R_{\rm A}\Omega$), one might take $v_{\rm A}\sim(v_\nu^2+R^2_{\rm A}\Omega^2)^{1/2}$
in eq.~(\ref{slowspin}). For comparison, 
Taam \& Spruit (1989) employ $v_{\rm A}\sim[v_\nu^2+(8/27)R^2_{\rm A}\Omega^2]^{1/2}$.} 

There are several features of winds from rapidly rotating magnetically-dominated 
PNSs that are very different from their NRNM counterparts.  Specifically, for the estimate given 
for $R_{\rm A}$ in eq.~(\ref{slowspin2}) to apply, $R_{\rm A}$ must be greater than $R_{\rm sonic}$,
the sonic point, where the radial velocity is equal to the local adiabatic sound speed ($c_s$).  
However, typical
NRNM models yield $R_{\rm sonic}$ in the range of hundreds of kilometers, much larger
than $R_{\rm A}$ as implied by eqs.~(\ref{eta2spin}) and (\ref{eta3spin}).
This apparent problem is reconciled easily; in highly magnetic, rapidly rotating PNS winds,
$R_{\rm sonic}$ occurs at radii much smaller than those inferred from NRNM models.
The boundary conditions at the PNS surface -- that the neutrinos are in thermal and chemical 
equilibrium with the matter and that the neutrino optical depth is $\sim2/3$ (see Thompson et al.~2001) --
conspire to give a sound speed ($c_s$) at $R_\nu$ of $\sim3\times10^9$ cm s$^{-1}$.  
Therefore, for $R_\nu=10$ km and spin period shorter than $\sim2$ ms ($\Omega\gtrsim3000$ rad s$^{-1}$)
$R_\nu\Omega$ exceeds $c_s(R_\nu)$.  This limit, $v_\phi(R_\nu)\gg c_s(R_\nu)$,
is termed `centrifugal' (Lamers \& Cassinelli 1999).  
In this limit, the centrifugal force dominates wind driving and we may neglect 
both neutrino heating and the gas pressure in the critical wind equations near $R_{\rm sonic}$ 
(see e.g.~eq.~\ref{vwind}; discussion of  the general wind equations is deferred to \S\ref{section:nmcwind}).
A simple expression results;
\beq
\left.v_e^2/2\right|_{R_{\rm sonic}}=\left.v_\phi^2\right|_{R_{\rm sonic}},
\eeq
where $v_e$ is the local escape velocity.
The location of the sonic point is then 
\beqa 
R_{\rm sonic}=(GM/\Omega^2)^{1/3}\,&=&\,(R_{\rm Sch}R_{\rm L}^2/2)^{1/3} \nonumber \\
\,\,&\simeq&\,\, 16.8\,\,{\rm km}\,\,\,M_{1.4}^{1/3}\,P_1^{-2/3},
\label{sonic}
\eeqa
where  $R_{\rm Sch}$ is the Schwarzschild radius ($\sim4.15$\,\,km for a 1.4 M$_\odot$ PNS).
With this estimate we find that the sonic point occurs  much closer to the
PNS and that $R_{\rm A}>R_{\rm sonic}$ for eqs.~(\ref{eta2spin}) and (\ref{eta3spin}).

There is another feature of centrifugal winds that differs significantly from NRNM outflows.
As $R_\nu\Omega$ becomes greater than $c_s(R_\nu)$  the 
matter density scale height increases exponentially with $\Omega^2$  due to centrifugal support
in the near-hydrostatic region ($v_r\ll c_s$).  For this reason, the derived mass loss rate also
increases exponentially with $\Omega^2$.  Note that this is only true if the magnetic
field supports the flow from the surface of the star to the sonic point.   That is,
we only expect large increases in $\dot{M}$ for $R_{\rm A}\ge R_{\rm sonic}$.
Examining eq.~(\ref{spintime})  we see that
with $\dot{M}$ increasing exponentially with $\Omega^2$, $\tau_{\rm J}$ may become
very short for large $\Omega$.
However, as $\Omega$ is increased and the mass
loading on the field lines increases, $B_0$ must be large enough to guarantee 
$R_{\rm A}>R_{\rm sonic}$ (eq.~\ref{slowspin2}).
In NRNM models, characteristic mass loss rates are less than
$10^{-4}$ M$_\odot$ s$^{-1}$.  The exponential effect on $\dot{M}$ in the centrifugal limit justifies our
use of a much higher mass loss rate in eqs.~(\ref{eta2spin}) and (\ref{eta3spin}), when $P=1$ ms
(see Fig.~\ref{plot:mdotthis} and \S\ref{section:nmcwind}).

 In an effort to make contact with the pulsar and solar wind communities, it is
 worth noting that taking eq.~(\ref{slowspin2})
 and multiplying both sides by $\Omega^3$ for $\eta=2$, one obtains
 \beq
 v_{\rm A}^3=v_{\rm M}^3\equiv\frac{\Omega^2 B_0^2R_\nu^4}{\dot{M}}=\frac{\Omega^2 F_{\rm B}}{F_{\rm M}},
 \label{vafast}
 \eeq
 where $F_{\rm B}$ is the magnetic flux, $F_{\rm M}$ is the matter flux, and $v_{\rm M}$ is
 the Michel velocity (Michel 1969; Belcher \& Macgregor 1976).  
 As one would expect from the Bernoulli integral,
 it is the Michel velocity that, to constants of order unity, is the asymptotic wind velocity.
 In a true fast ($v_\nu\ll R_{\rm A}\Omega$), magnetically dominated wind the Alfv\'{e}n velocity is 2/3 of the Michel
 velocity, and $v_{\rm M}$ obtains only at the fast magnetosonic point (e.g. Belcher \& MacGregor 1976).
 In fact, for a strictly monopolar cold ($c_s=0$ everywhere)  magnetocentrifugal wind the
 fast point is formally at infinity (Michel 1969).
 At the level of accuracy aspired to here, these details are unimportant.

Spindown timescales ($\tau_{\rm J}=\Omega/\dot{\Omega}$; eq.~\ref{spintime}) can be computed from the
estimates of $R_{\rm A}$ in eqs.~(\ref{eta2spin}) and (\ref{eta3spin}).  
Together with eq.~(\ref{slowspin2}) we find that
\beq
\tau^{\rm NR}_{\rm J}=
(2/5)M\,\dot{M}^{(3-2\eta)/(2\eta-1)}\,R_\nu^{-2/(2\eta-1)}\,
B_0^{-4/(2\eta-1)}\,\Omega^{\,2/(2\eta-1)}.
\label{timescaleeta}
\eeq
The superscript `NR' is added to emphasize that when the flow is non-relativistic,
$\tau_{\rm J}$ depends explicitly on $\dot{M}$ as in eq.~(\ref{spintime}).  For relativistic
flows it does not (see eq.~\ref{timescaleetarel}). Taking $M=1.4$\,\,M$_\odot$ and scaling,
\beqa
\tau^{\rm NR}_{\rm J}(\eta=2) &\simeq& 30\,\,{\rm s}\,\,\,
\mns\,\mdotth^{-1/3}\,\rnuten^{-2/3}\,\bof^{-4/3}\,P_{1}^{-2/3}
\label{slowspinscale2} \\
\tau^{\rm NR}_{\rm J}(\eta=3)&\simeq&96\,\,{\rm s}\,\,\,
\mns\,\mdotth^{-3/5}\,\rnuten^{-2/5}\,\bof^{-4/5}\,P_{1}^{-2/5}
\label{slowspinscale3}
\eeqa
The important timescale to compare with $\tau_{\rm J}^{\rm NR}$ is the 
Kelvin-Helmholtz cooling timescale, $\tau_{\rm KH}$.  This comparison is relevant because $\dot{M}$
drops steeply with the PNS neutrino luminosity.  If $\tau_{\rm J}^{\rm NR}\gg\tau_{\rm KH}$
no spindown occurs during the wind epoch.  
The PNS cooling calculations of Pons et al.~(1999) show that $L_\nu^{\rm tot}$ drops by a factor
of ten, from $\sim10^{52}$ erg s$^{-1}$ to $\sim10^{51}$ erg s$^{-1}$, in $\sim30$ seconds.  This result
depends on the mass of the PNS and the high-density nuclear equation of state employed.  
These calculations do not include the effects of rapid rotation or high magnetic fields.  Both 
may significantly modify $\tau_{\rm KH}$.  Rapid rotation leads to lower core temperatures and,
thereby, lower average neutrino luminosity. This could increase $\tau_{\rm KH}$ significantly
(see Thompson, Quataert, \& Burrows, in prep; Yuan \& Heyl 2003; Villain et al.~2003).  Rapid rotation
might also make $R_\nu$
contract more slowly, affecting both $B_0$ and $\Omega$ at a given time because both are
proportional to $R_\nu^{-2}$, via flux and angular momentum conservation, respectively.
High $B_0$ may also affect $\tau_{\rm KH}$ by modifying neutrino opacities (e.g.~Lai \& Qian 1998;
Arras \& Lai 1999).  Multi-dimensional effects such as convection may also be important
(Keil, Janka, \& M\"{u}ller 1996).  Without detailed models of magnetar cooling, we take
the results of Pons et al.~(1999) as representative and define $\tau_{\rm KH}\sim30$ seconds.

As argued in \S\ref{section:eta} the spindown timescales relative to $\tau_{\rm KH}$
are sensitive both in absolute value and scaling to the parameter $\eta$.
Even allowing for our ignorance about the magnetic field topology,
eqs.~(\ref{slowspinscale2})  and (\ref{slowspinscale3}) 
imply that at best $\tau_{\rm J}^{\rm NR}\sim\tau_{\rm KH}$ for the PNS parameters chosen.  
Note that for higher $\dot{M}$, both timescales
decrease somewhat, despite the fact that $R_{\rm A}$ decreases as $\dot{M}$ increases.
Interestingly, for larger initial spin periods, the spindown timescales {\it decrease}.
Hence, if magnetars are born with $P\sim30$ ms -- consistent with the spin periods of most pulsars today --
then $\tau_{\rm J}^{\rm NR}\sim 3$ s and $\sim25$ s for $\eta=2$ and $\eta=3$, respectively.
As we will see in \S\ref{section:nmcwind} mass loss rates as high as $10^{-3}$ M$_\odot$ s$^{-1}$
are probably unrealistic at such low $\Omega$.  Even so, spindown of slowly rotating 
magnetars during the cooling epoch can be quite efficient.
For higher $B_0$, both spindown timescales drop rapidly. Increasing the surface magnetic field
strength to $10^{16}$ G in eq.~(\ref{slowspinscale3}), $\tau^{\rm NR}_{\rm J}(\eta=3)\simeq15$ s,
which may be considerably less than $\tau_{\rm KH}$.
Although such a large $B_0$ is not out of the question in this very early
phase of PNS evolution, it is well beyond the surface field strengths inferred from
magnetars (Kouveliotou et al.~1999; Thompson \& Duncan 1993).

Consideration of such high field strengths brings up an important physical constraint on $R_{\rm A}$,
particularly germane for PNSs rotating rapidly at birth:
$R_{\rm A}$ cannot be made arbitrarily large by increasing $B_0$ for a given $\Omega$ because
$v_\phi$ would eventually exceed the speed of light.
Thus, the Alfv\'{e}n point, determined {\it a posteriori} for a given $\Omega$, must be less than the light 
cylinder radius, $R_{\rm L}=c/\Omega$,
although  for asymptotically large ratios of magnetic flux to matter flux $R_{\rm A}$ asymptotes to $R_{\rm L}$.  
This requirement sets a critical $B_0$.  If we set $R_{\rm A}=R_{\rm L}=c/\Omega$ 
in eq.~(\ref{slowspin2}) for a given $\Omega$, $R_\nu$, and $\dot{M}$, the $B_0$ derived
is that required to enforce co-rotation to {\it approximately} $R_{\rm L}$.  
This critical magnetic field, for arbitrary $\eta$, is easily derived in terms of the basic wind and PNS parameters;
 \beq
 B_{\rm crit}=c^{(2\eta-1)/2}\,\dot{M}^{1/2}\,\Omega^{1-\eta}\,R_\nu^{-\eta}.
 \label{bcrit}
 \eeq
For example, for a neutron star with spin period of 1 ms, $R_{\rm L}\simeq47.7$ km
and for $\dot{M}=10^{-3}$ M$_\odot$ s$^{-1}$ as in eqs.~(\ref{slowspinscale2}) and (\ref{slowspinscale3}),
$B_{\rm crit}\simeq1.2\times10^{15}$ G and $5.7\times10^{15}$ G for $\eta=2$ and $\eta=3$, respectively.

If $R_{\rm A}$ is sufficiently less than $R_{\rm L}$, the asymptotic velocity of the wind material
is not relativistic.  In non-relativistic outflows, consideration of the Bernoulli integral
in the Weber-Davis model shows that the energy carried by the magnetic field 
is a factor of two greater than that of the matter (e.g.~Lamers \& Cassinelli 1999) and the 
rotational energy loss rate is 
\beq
\dot{E}_{\rm tot}=I\Omega\dot{\Omega}=-\dot{M}{\cal L}\Omega=-B_0^{4\zeta}R_\nu^{4\eta\zeta}\dot{M}^{1-2\zeta}\Omega^{2-2\zeta},
\label{nonreletot}
\eeq
where $\zeta=1/(2\eta-1)$.
If $B_0$ exceeds $B_{\rm crit}$, $R_{\rm A}$ approaches $R_{\rm L}$ and $v_{\rm A}$ approaches $c$.
The flow becomes relativistic and the field may carry more than two times the energy of the matter.  
At $R_{\rm L}$, the ratio of magnetic flux to matter flux is given by\footnote{The ratio $\Gamma$, as defined in eq.~(\ref{gammalim}), is often denoted by `$\sigma$' in the pulsar literature.}  
\beq
\Gamma=\left. \frac{B^2}{4\pi\rho c^2}\right|_{R_{\rm L}}=\,\,\,B_0^2 R_\nu^{2\eta}\Omega^{2\eta-2}c^{1-2\eta}\dot{M}^{-1}.
\label{gammalim}
 \eeq
and the relativistic energy loss rate is
\beq
\dot{E}_{\rm tot}=-\Gamma\dot{M}c^2= -B_0^2 R_\nu^{2\eta}\Omega^{2\eta-2}c^{3-2\eta}.
\label{reletot}
\eeq 
Note that $\dot{E}_{\rm tot}$ is independent of $\dot{M}$, but depends on the magnetic field structure.
For the case $\eta=3$, the pure magnetic vacuum dipole scaling for $\dot{E}_{\rm tot}$ is obtained.\footnote{Note 
that in the standard theory
(see e.g.~Shapiro \& Teukolsky 1983 chapter 10 for a review) $\dot{E}_{\rm tot}$ in the vacuum dipole limit
includes a factor $\sin^2\alpha/6$, where $\alpha$ is the angle between the spin axis of
the neutron star and its magnetic axis.}
For $\eta=2$, this expression for $\dot{E}_{\rm tot}$ implies a much larger energy loss
rate than for $\eta=3$ -- the ratio being $c^2/R_\nu^2\Omega^2$, a factor of $\sim23$ for a 10 km PNS
with a 1 ms spin period.  As $\dot{M}$ drops during the cooling epoch, for constant $B_0$, the wind
transitions from a non-relativistic (eq.~\ref{nonreletot}) to a relativistic outflow  (eq.~\ref{reletot}).
Only in the case where $\Omega$ is very small so that $R_{\rm L}$ is very large are there instances
when the wind is always non-relativistic during $\tau_{\rm KH}$ (see \S\ref{section:extract}).
Taking eq.~(\ref{gammalim}), the spindown timescale ($\tau_{\rm J}=\Omega/\dot{\Omega}$) 
for relativistic winds, in analogy with eq.~(\ref{timescaleeta}), is
\beq
\tau_{\rm J}^{\rm R}=(2/5)MR_\nu^{2-2\eta}B_0^{-2}\Omega^{4-2\eta}c^{2\eta-3}.
\label{timescaleetarel}
\eeq
Taking the same scalings as in eqs.~(\ref{slowspinscale2}) and (\ref{slowspinscale3})
\beqa
\tau^{\rm R}_{\rm J}(\eta=2) &\simeq& 34\,\,{\rm s}\,\,\,
\mns\,\rnuten^{-2}\,\bof^{-2}
\label{relspinscale2} \\
\tau^{\rm R}_{\rm J}(\eta=3)&\simeq&760\,\,{\rm s}\,\,\,
\mns\,\rnuten^{-4}\,\bof^{-2}\,P_{1}^{2}
\label{relspinscale3}
\eeqa
The superscript `R' is meant to reinforce the difference between relativistic and
non-relativistic wind spindown -- the latter depending explicitly on $\dot{M}$.
Comparing eqs.~(\ref{relspinscale3})  and (\ref{slowspinscale3}) we see that $\tau_{\rm J}^{\rm NR}$
is significantly shorter than $\tau_{\rm J}^{\rm R}$.   The difference is one of applicability.
Equation (\ref{slowspinscale3}) applies with large $\dot{M}$, when $R_{\rm A}<R_{\rm L}$ and $\Gamma<1$.
Equation (\ref{relspinscale3})  applies only when the mass flux from the PNS subsides to such an extent 
that $\Gamma\ge1$.  Taking $B_0=B_{\rm crit}$ in eq.~(\ref{slowspinscale3}), $\tau_{\rm J}^{\rm NR}=\tau_{\rm J}^{\rm R}$.
Therefore, application of vacuum dipole spindown is appropriate only when $\dot{M}$ is sufficiently
small.  We see that a naive and inappropriate application of eq.~(\ref{relspinscale3})  when the
mass flux is $\sim10^{-3}$ M$_\odot$ s$^{-1}$ under-predicts the rate of angular momentum loss by a factor
of $\sim8$.\footnote{Although both considered magnetars in the context of soft gamma repeaters and
anomalous X-ray pulsars, Thompson \& Blaes (1998)  and Harding
et al.~(1999) showed that enhanced spindown is obtained in models including relativistic particle winds.}
Note that this discrepancy is much larger for slowly rotating magnetars;
$\tau_{\rm J}^{\rm R}(\eta=3)/\tau_{\rm J}^{\rm NR}(\eta=3)\propto P^{12/5}$.  If magnetars are born
with $P=10$ ms, $\tau_{\rm J}^{\rm R}/\tau_{\rm J}^{\rm NR}\sim2000$ and one may under-predict spindown
by several orders of magnitude.  Of course, the difference is further magnified if, when the mass loss rate
is high, the field is monopolar to some extent, as argued in \S\ref{section:eta}.  In this
case we compare eq.~(\ref{relspinscale3}) ($\eta=3$)  with  eq.~(\ref{slowspinscale2}) ($\eta=2$) and find that 
the vacuum dipole limit overestimates the spindown timescale by a factor of $\sim25$ for the fiducial 
1 ms rotator with $B_0=10^{15}$ G.  Noting that $\tau_{\rm J}^{\rm R}(\eta=3)/\tau_{\rm J}^{\rm NR}(\eta=2)
\propto P^{8/3}$, we see that this ratio is $\sim10^4$ for a slow rotator with $P=10$ ms.

\section{Neutrino-Magnetocentrifugal Winds on the Cheap}
\label{section:nmcwind}

Because the spindown timescales and
the amount of rotational energy extractable depend so crucially on $\dot{M}$,
we solve here the steady-state wind equations including neutrino heating and rotation
explicitly.  Magnetic fields are included implicitly by enforcing $v_\phi=r\Omega$
everywhere. 
Thus, the wind mass elements are forced to corotate with the PNS surface as if 
on wires emanating radially from the equator of the PNS like the spokes 
on a bike wheel. This approximates the effect of a strong magnetic field or the 
non-relativistic Weber-Davis problem.  Of course, in our solution the mass elements
eventually have velocity greater than $c$.  For a given $\Omega$, the radius at which
this occurs is the light cylinder.  For this reason our solutions are
valid (if approximately) only for $r<R_{\rm L}$.  Importantly, since it is the
effect of the centrifugal force on $\dot{M}$ that we are most interested in, and
$\dot{M}$ will be most effected by the centrifugal force in the subsonic region,
the sonic point (see eq.~\ref{sonic}) is always well within $R_{\rm L}$ where the corotation
assumption is valid.

It would be preferable to solve the one-dimensional relativistic Weber-Davis problem (Michel 1969; Goldreich \& Julian 1970)
with an arbitrary energy deposition function.  We save this for a future paper.\footnote{
We have developed a one-dimensional non-relativistic time-dependent Eulerian 
magnetohydrodynamics code for solving the isothermal Weber-Davis problem.  Preliminary
comparisons between models in which strict corotation is enforced ($v_\phi=r\Omega$) everywhere
and the actual Weber-Davis solution show that  $R_{\rm A}$ is 
typically $\sim1.5$ times {\it larger} than that expected from the models presented
here.  This implies that we underestimate the magnitude of the spindown for a given $B_0$
and overestimate spindown timescales $\tau_{\rm J}^{\rm NR}$ by a factor of $\sim2$.}
Much more important,
we feel, is our assumption that the problem may be treated in one spatial dimension 
(as in Weber \& Davis 1967) in the presence of rapid rotation and strong magnetic fields (see \S\ref{section:eta}).  
This approximation, can only be
removed with a code capable of two-dimensional axisymmetric relativistic MHD for $\beta\ll1$.
This is a considerable effort beyond the scope of the current work.

\subsection{Equations}

Enforcing continuity, conservation of momentum, and conservation of energy
in Newtonian gravity,
we have that
\beq
\frac{1}{v_r}\frac{dv_r}{dr}\,(c_s^2-v_r^2)=\frac{2}{r}\left(\frac{v_e^2}{4}-\frac{v_\phi^2}{2}-c_s^2\right)+
\dot{q}\frac{D}{C_VTv_r}
\label{vwind}
\eeq
\beq
\frac{1}{\rho}\frac{d\rho}{dr}\,(c_s^2-v_r^2)=\frac{2}{r}\left(v_r^2-\frac{v_e^2}{4}+\frac{v_\phi^2}{2}\right)-
\dot{q}\frac{D}{C_VTv_r}
\label{rhowind}
\eeq
\beq
\frac{D}{T}\frac{dT}{dr}\,\left(\frac{c_s^2-v_r^2}{c_s^2-c_T^2}\right)=\frac{2}{r}
\left(v_r^2-\frac{v_e^2}{4}+\frac{v_\phi^2}{2}\right)
+\dot{q}\frac{D}{C_VTv_r}\left(\frac{c_T^2-v^2}{c_s^2-c_T^2}\right),
\label{twind}
\eeq
where $v_e=(2GM/r)^{1/2}$, $D=(T/\rho)\p P/\p T|_\rho$, $c_T$ and $c_s$ are
the isothermal and adiabatic sound speeds, respectively, $C_V$ is the specific heat at constant volume,
$v_\phi=r\Omega$, $v_r$ is the radial velocity, and $\dot{q}$ is the specific neutrino heating rate.  
For a given $\Omega$, we solve the above equations
for the flow between the surface of the PNS and the sonic point as a two-point boundary value problem,
using a relaxation algorithm, on an adaptive radial mesh (see Thompson et al. 2001).
We integrate from the sonic point to the light cylinder using a simple Runga-Kutta algorithm.
The equation of state (EOS) assumes ideal nucleons.  It includes a general electron-positron
EOS, photons, and neglects the formation of alpha particles.  Magnetic effects on the
EOS, particularly important for electron-positron component at low temperatures, have not been included.
Once a solution is obtained, we can see what surface magnetic field strength is required to
corotate out to any radius.  Once the field is chosen, the solution is valid only inside this
{\it a posteriori}-posited Alfv\'{e}n point.  Outside this radius, our solutions continue to accelerate
as a result of the centrifugal force.  In the full Weber-Davis problem, $v_\phi$ begins
to decrease at $R_{\rm A}$ and $v_r$ asymptotes, reaching the Michel velocity at the fast magnetosonic 
point (see Belcher \& Macgregor 1976).  

\subsection{NMC Wind Results}

\begin{figure}
\begin{center}
\includegraphics[height=15cm,width=15cm]{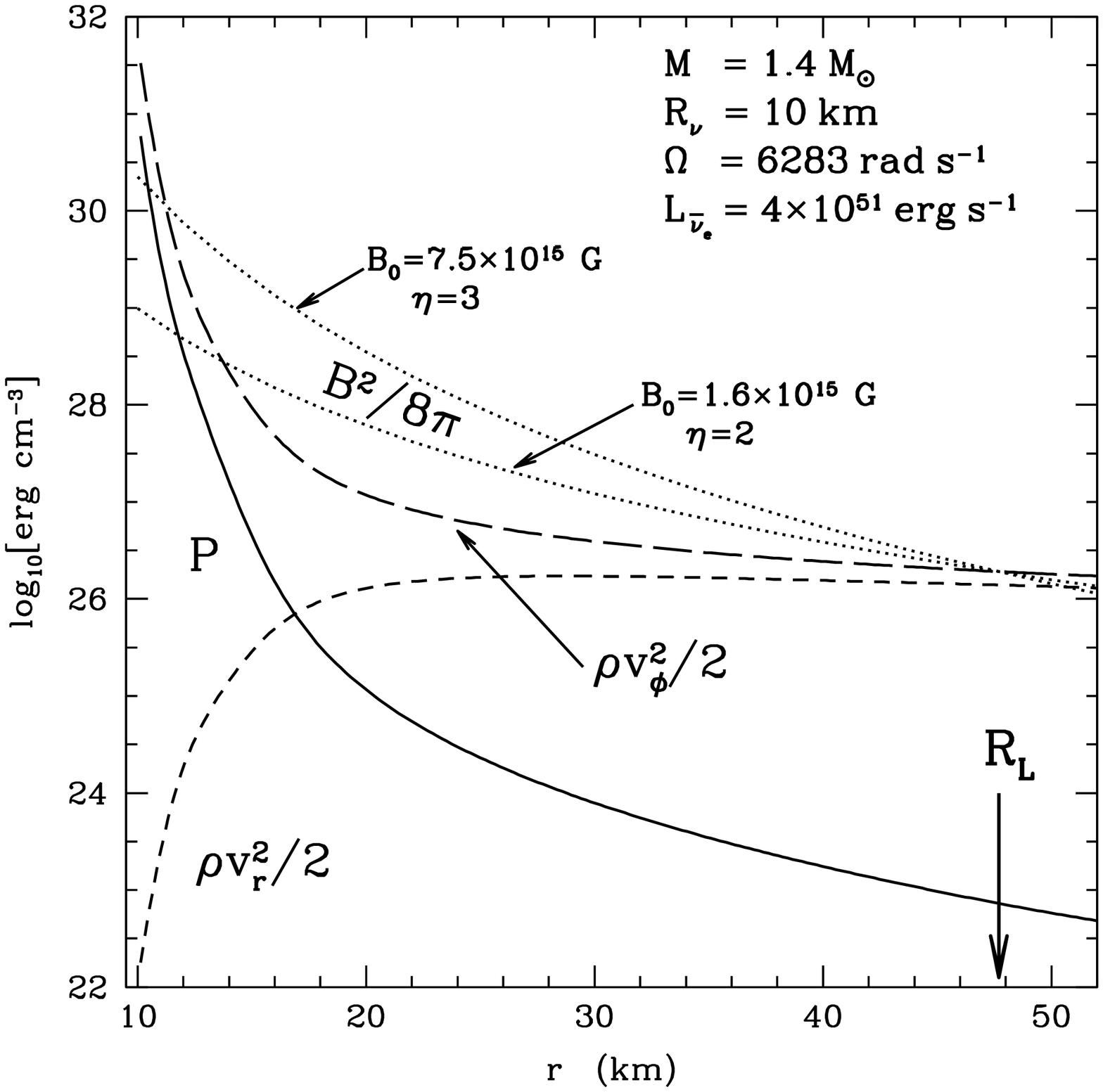}
\end{center}
\caption{Thermal pressure ($P$, solid line), kinetic energy density ($\rho v_r^2/2$, short dashed line;
$\rho v_\phi^2/2$, long dashed line), and magnetic energy density ($B^2/8\pi$, dotted lines) for $B_0=1.6\times10^{15}$\,G and $B_0=7.5\times10^{15}$\,G for $\eta=2$ and $\eta=3$, respectively, as a function of radius.
The PNS has spin period of 1\,\,ms ($\Omega\simeq6283$ rad s$^{-1}$, $R_L\simeq47.7$\,km is 
the corresponding  light cylinder radius),
$L_{\bar{\nu}_e}=4\times10^{51}$ erg, $R_\nu=10$ km, and $M=1.4$ M$_\odot$.   
The mass loss rate is $\dot{M}\simeq1.6\times10^{-3}$ M$_\odot$ s$^{-1}$
(compare with Fig.~\ref{plot:mdotthis}).  Model numbers were chosen to closely follow those employed for the
scaling relations in \S\ref{section:analytic}.}
\label{plot:pp}
\end{figure}

Figure \ref{plot:pp}  shows the thermal pressure (solid line) and kinetic energy density
($\rho v_\phi^2/2$ long dashed line and $\rho v_r^2/2$ short dashed line) 
for a PNS with $L_{\bar{\nu}_e}=4\times10^{51}$ erg s$^{-1}$,
$M=1.4$ M$_\odot$, and $R_\nu=10$ km.  
We use $L_{\bar{\nu}_e}$ to label wind models; the total neutrino luminosity for a given model is $L_\nu^{\rm tot}=
L_{\nu_e}+L_{\bar{\nu}_e}+4L_{\nu_\mu}$, with $L_{\nu_e}=L_{\bar{\nu}_e}/1.3$ and
$L_{\nu_\mu}=L_{\bar{\nu}_e}/1.4$.  The subscript `$\nu_\mu$' stands for $\nu_\mu$,
$\bar{\nu}_\mu$, $\nu_\tau$, and $\bar{\nu}_\tau$.  
For $L_{\bar{\nu}_e}=4\times10^{51}$ erg s$^{-1}$, 
$L_\nu^{\rm tot}\simeq1.85\times10^{52}$ erg s$^{-1}$.
The PNS period is taken to be 1\,ms ($\Omega=6283$ rad s$^{-1}$).
Overlayed on Fig.~\ref{plot:pp} is the magnetic energy
density (dotted lines, labeled $B^2/8\pi$) for $\eta=2$ and $\eta=3$ such that $B^2/8\pi\sim\rho c^2/2$ at 
$R_{\rm L}$.  The point at which  the dotted curves intersect the short dashed lines corresponds
to the Alfv\'{e}n radius.  The surface magnetic field strengths are labeled $B_0$ and correspond to $B_{\rm crit}$
in eq.~(\ref{bcrit}) for $\eta=2$ and $\eta=3$.  For these field strengths, then, the Alfv\'{e}n point corresponds
to $\sim R_{\rm L}$.  For lower $B_0$, $R_{\rm A}$ decreases. 
Figure \ref{plot:pp} shows that  we can expect a $1.6\times10^{15}$ G surface magnetic
field with monopolar field topology to enforce corotation out to $\sim R_{\rm L}$  for mass loss rates in the 
range of $10^{-3}$ M$_\odot$ s$^{-1}$ (the model shown has $\dot{M}\simeq1.6\times10^{-3}$ M$_\odot$ s$^{-1}$).  
For $\eta=3$, the required surface field strength is $\sim7.5\times10^{15}$ G.
As $L_\nu^{\rm tot}$ increases $\dot{M}$ increases and larger $B_0$  is required to force corotation out to $R_{\rm L}$.
As $L_\nu^{\rm tot}$ decreases, the converse is true.
For example, for $L_{\bar{\nu}_e}=0.5\times10^{51}$ erg s$^{-1}$,
the required surface field strengths are $1.4\times10^{15}$\,G and $2.9\times10^{14}$\,G for
$\eta=3$ and $\eta=2$, respectively.  That $B_{\rm crit}$ decreases as $L^{\rm tot}_\nu$ decreases
is expected from eq.~(\ref{bcrit}) and the
fact that the mass loading of the field lines, $\dot{M}$, decreases steeply with $L_\nu^{\rm tot}$.

The intersection of $P$ with $\rho v_r^2/2$ (at $r\sim17$ km) marks the radius of the sonic point
and is very well described by eq.~(\ref{sonic}) for short PNS spin periods, regardless of $L_\nu^{\rm tot}$.
Note that for the $B_0$ used to plot $B^2/8\pi$ in Fig.~\ref{plot:pp}, for both $\eta=2$ and $\eta=3$,
that deep in the exponential atmosphere of the PNS, both $P$ and $\rho v_\phi^2/2$ become greater
than the magnetic energy density.   This is potentially important since we have assumed in constructing 
these models that corotation obtains ($v_\phi=r\Omega$) everywhere.  
This assumption is most important between $R_\nu$ and $R_{\rm sonic}$,
because centrifugal support in this region can increase $\dot{M}$ significantly.
The surface 
of the PNS calculations, $R_\nu$, is defined as the neutrinosphere (optical depth $\sim2/3$).  This
condition determines $\rho(R_\nu)$, which, for most calculations is $\sim10^{12}$ g cm$^{-3}$.
The azimuthal kinetic energy density at $R_\nu$ is then determined by setting $\Omega(R_\nu)$.
Although in Fig.~\ref{plot:pp} and \S\ref{section:analytic} we have assumed that $R_\nu$ 
corresponds to the radius at which the magnetic field is anchored ($B(r)=B_0(R_\nu/r)^\eta$),
this need not hold necessarily.  In addition, very close to the PNS, where the radial flow is
very subsonic, we expect the field to be a complex of higher-order multi-poles as shear
energy within the PNS convective region emerges as magnetic flux into the magnetically-dominated atmosphere
via the Parker instability (Thompson \& Murray 2001).  The degree of magnetic support within this
atmosphere, particularly if differentially rotating, will depend on the dynamics of the underlying
convection and the efficacy of the dynamo mechanism, issues that require a multi-dimensional
model of these MHD winds.

The importance of the combined action of rapid rotation and strong magnetic fields 
is seen most clearly in Figure \ref{plot:mdotthis}.
Here, we present the mass loss rate $\dot{M}$ for models with 
$L_{\bar{\nu}_e}=8\times10^{51}$ erg s$^{-1}$  and $0.5\times10^{51}$ erg s$^{-1}$ 
as a function of $\Omega$.  This range of neutrino luminosity corresponds to roughly 20 seconds
of the cooling epoch in the models of Pons et al.~(1999).
The solid lines assume strict corotation with the PNS surface, $v_\phi=r\Omega$.
The dashed lines are presented for comparison assuming no corotation and angular momentum conservation;
in these models $v_\phi=R_\nu\Omega(R_\nu/r)$. 
All mass loss rates quoted are `spherical' and assume $\dot{M}=4\pi r^2\rho v_r$.
For $\Omega>3000$ rad s$^{-1}$ in the models assuming corotation, significant deviations from the
$\Omega=0$ limit are obtained.  This results from the fact that at this $\Omega$, for $R_\nu=10$ km, 
$R_\nu\Omega$ becomes greater than $c_s(R_\nu)$.  As alluded to previously (\S\ref{section:analytic}), this changes the
mass flux in a dramatic way; for $\Omega=6283$ rad s$^{-1}$ and $L_{\bar{\nu}_e}=8\times10^{51}$ erg s$^{-1}$, $\dot{M}\simeq8\times10^{-3}$
M$_\odot$ s$^{-1}$, more than 30 times larger than in the non-rotating model
($\dot{M}\simeq2.6\times10^{-4}$ M$_\odot$ s$^{-1}$).  For comparison, with 
$\Omega=9000$ rad s$^{-1}$, $\dot{M}\simeq0.36$
M$_\odot$ s$^{-1}$.  
For $\Omega\gg3000$ rad s$^{-1}$, the functional dependence of $\dot{M}$ can be approximated
roughly with 
\vspace*{.1cm}
\beq
\dot{M}(\Omega R_\nu\gg c_s(R_\nu))\propto \exp\left[\Omega^2R_\nu^2/c_s^2(R_\nu)\right]
\label{mdotomega}
\eeq
as expected from simple considerations of the subsonic exponential atmosphere of the PNS.
For $\Omega\sim3000$ rad s$^{-1}$, $\dot{M}$ depends on the location of the sonic point itself
(see e.g.~Belcher \& Macgregor 1976; Hartmann \& Macgregor 1982; Lamers \& Cassinelli 1999).  
For rapidly rotating PNSs the spindown timescale may be very short as a result of the large
increase in $\dot{M}$ expected when $R_\nu\Omega\gg c_s$ (eqs.~\ref{slowspinscale2} and \ref{slowspinscale3}).  
As an example, taking 
$L_{\bar{\nu}_e}=8\times10^{51}$ erg s$^{-1}$, with $P=1$ ms,
$\eta=2$, and $B_0\sim B_{\rm crit}\sim3.6\times10^{15}$ G,
we have from eq.~(\ref{slowspinscale2}) that $\tau_{\rm J}^{\rm R}\sim\tau_{\rm J}^{\rm NR}\sim3.1$ seconds.
These numbers, together with the scaling relations for $\tau_{\rm J}^{\rm NR}$ and $\tau_{\rm J}^{\rm R}$ 
in \S\ref{section:analytic}, 
imply that a significant amount of rotational energy may be
extracted from the PNS during the neutrino cooling epoch.

\vspace*{1cm}

\subsection{Rotational Energy Extraction \& Spindown}
\label{section:extract}

The spindown timescale depends crucially on $\eta$.  This is seen in eqs.~(\ref{timescaleeta})
and (\ref{timescaleetarel}).   
In addition to the arguments of \S\ref{section:eta}, there are a few things that can be argued with
some certainty about the field structure.  
First, if $B^2/8\pi\ll P$ at the surface, then no closed magnetic field lines will persist and $B_r\propto r^{-2}$,
despite the fact that the currents in the star should necessarily produce a dipole field.
Although uninteresting from the perspective of spindown because no $R_{\rm A}$ exists outside 
$R_\nu$, certainly in this limit we should have $\eta\sim2$.  
Second, and at the other extreme, in a totally magnetically dominated system with
$\Gamma$ very large, we expect the field to be dipolar with closed field lines within $R_{\rm L}$ (although, see
\S\ref{section:eta}).
In this limit, if we accept the magnetic dipole spindown model for magnetars (and neutron stars
generally), then we must have $\eta\sim3$.  These simple arguments suggest that $\eta$ itself is a function
of the ratio of the magnetic flux to the matter flux.  Following Ud-Doula \& Owocki (2002), 
one may construct the dimensionless number 
\beq
\xi=B_0^2R_\nu^2/\dot{M}v_\infty
\eeq
and write $\eta$ as a function of $\xi$.  However, with the complications of rapid rotation and
relativity, we have decided to simply bracket the possible solution space, computing solutions
with $\eta=2$ and $\eta=3$ separately as in \S\ref{section:analytic}.

{\bf Monopole Spindown:}  We solve the differential equation $I\Omega\dot{\Omega}=\dot{E}_{\rm tot}$ 
for $\Omega(t)$ with $\eta=2$.  If $R_{\rm A}<R_{\rm L}$ (non-relativistic wind), we use $\dot{E}_{\rm tot}=-\dot{M}R_{\rm A}^2\Omega^2$,
with $R_{\rm A}$ from eq.~(\ref{slowspin2}) with $\eta=2$.  If $\Gamma>1$ (relativistic wind), we take $\dot{E}_{\rm tot}=\Gamma\dot{M}c^2$,
with $\Gamma=B_0^2 R_\nu^{4}\Omega^{2}\dot{M}^{-1}c^{-3}$.  In order to obtain $\dot{M}$, we employ
the results from our NMC models.
We assume that these
steady-state wind solutions, for different PNS characteristics, may be concatenated to form a time series.
The total neutrino heating rate is computed self-consistently, assuming a PNS neutrino luminosity,
including contributions from $\nu_e n\leftrightarrow p e^-$, $\bar{\nu}_e p\leftrightarrow n e^+$,
$\nu\bar{\nu}\leftrightarrow e^+ e^-$, and inelastic scattering of neutrinos with charged leptons
and nucleons (Thompson et al.~2001).  For the models presented here, as reflected by the
requirement that $R_{\rm A}\Omega$ is always greater than $v_\nu$ (see \S\ref{section:analytic}), the total energy deposition
rate via neutrino heating and cooling is more than an order of magnitude less $\dot{M}v_\infty^2$.
This implies that all wind models are primary driven magnetocentrifugally.
In NRNM models of PNS winds, if $\varepsilon_\nu\propto L_\nu^{1/4}$, where $\varepsilon_\nu$
is the average neutrino energy, $\dot{M}\propto L_\nu^{5/2}$ (Qian \& Woosley 1996).\footnote{This 
dependence of the mass loss rate on luminosity is only approximate.
For very large $\Omega$ (above $7000$ rad s$^{-1}$) this power law dependence of $\dot{M}$
on $L_\nu$ is modified and the exponent decreases.}  
With this dependence in hand from our wind 
models we need only posit $L_\nu(t)$.
Taking $L_\nu(t<40\,\,{\rm s})\propto t^{-\alpha}$,
with $\alpha$ in the range $\sim1$ as appropriate over much of the cooling epoch (see Pons et al.~1999),
and $L_\nu(t>40\,\,{\rm s})\propto e^{-t/\tau}$ with $\tau=1$ second, we can compute $\Omega(t)$.  

As an example of slow magnetic rotator spindown we take an initial spin period of 20 ms and 
$B_0=1.5\times10^{15}$ G. Integrating from $t_i=1$ s to $t_f=100$ s, we find that the PNS stops 
rotating in $\sim50$ seconds.  In this model a total of $1.75\times10^{-4}$ M$_\odot$ is ejected
and all of the rotational energy of the PNS is extracted.  For this
choice of parameters, $\Gamma<1$ throughout the evolution. For longer spin periods 
we find similar evolution. The PNS simply stops rotating in tens of seconds.
Hence, if magnetars are born slowly rotating (with an initial spin period similar to normal neutron stars)
we find that monopolar neutrino-magnetocentrifugal
winds can efficiently extract the rotational energy of the young PNS on the Kelvin-Helmholtz timescale.

As an example of an evolutionary sequence in which we start with a rapidly rotating PNS,
we take the initial spin period to be 1 ms and the same temporal evolution
of the neutrino luminosity, folding in the exponential dependence of $\dot{M}$ on $\Omega$
for $\Omega>3000$ rad s$^{-1}$ (as in Fig.~\ref{plot:mdotthis}).  
Results for $B_0=3\times10^{15}$ G (A), $1.5\times10^{15}$ G (B), $10^{15}$ G (C), and $7\times10^{14}$ G (D) 
are presented in Figure \ref{plot:allp}. The posited time evolution of  $L_\nu$ 
is shown in the lower right panel.  The evolution of the mass loss rate follows from the models
constructed in \S\ref{section:nmcwind} and is shown in the upper right panel.
Computed evolution of $P(t)$ is in the upper left panel.  For clarity of presentation,
only the first $\sim20$ seconds of $P(t)$ for trajectory (A) are shown.
The lower left panel shows the total rotational energy extracted from the PNS
as a function of time (solid lines, $10^{51}$ erg) and the rotational 
energy loss rate (dashed lines, $10^{51}$ erg s$^{-1}$).  The lower middle panel shows $\Gamma$
as a function of $t$ for models (A)--(D), and a model with $B_0=1.3\times10^{15}$ G (heavy dotted line).
Finally, the upper middle
panel shows the time evolution of $R_{\rm L}$ (solid lines) and $R_{\rm A}$ with $\eta=2$ (dashed lines).

In model (A) the flow begins with $R_{\rm A}\sim R_{\rm L}$.  For model (D) several 
seconds pass before the flow becomes relativistic and $\Gamma$ becomes greater than 1.
For all models shown here, over the 100 seconds of evolution computed, a total of $\gtrsim2\times10^{52}$ erg 
in rotational energy is lost by the PNS.  
Because of our assumption of $\eta=2$, the PNS spin period continues to $e$-fold 
on the timescale given by eq.~(\ref{relspinscale2}).
The total mass lost by the PNS decreases as $B_0$ increases;  for model (A)
$3.3\times10^{-3}$ M$_\odot$ is ejected, whereas for model (D) $5.3\times10^{-3}$ M$_\odot$ is emitted.
Because $\Omega$ drops so rapidly in model (A), and $\dot{M}$ depends exponentially on $\Omega$
for $P\lesssim2$ ms, this correlation is expected.
In models (B), (C), and (D), more than $10^{52}$ erg is emitted with $\Gamma>100$.  In fact, at the end of these spindown
calculations, with $\dot{M}$ plummeting with the exponential cutoff in $L_\nu$, $\Gamma$
can easily reach 1000.  The exact evolution depends sensitively on $B_0$ and the time evolution of $\dot{M}$,
as can be seen from the model with $B_0=1.3\times10^{15}$ G (dotted line) in the lower middle panel.
The complex time evolution of $\Gamma$ comes both from the exponential
cut-off in $\dot{M}$ (imposed at $t=40$ s) and the fact that $\Gamma$ can decrease
even as the mass loss abates if $R_{\rm L}$ increases rapidly.  That is, for constant $B_0$,
$\Gamma$ can decrease if the spindown timescale is shorter than the timescale for $\dot{M}$ to decrease (see eq.~\ref{gammalim}).
The non-relativistic epoch of spindown ($\Gamma<1$) lasts a mere $\sim1-2$ s in models (B)--(D),
and accounts for $\sim2-3\times10^{51}$ erg in rotational energy extracted.
Although this timescale is short on the scale of $\tau_{\rm KH}$, it is very long on the
scale of the PNS dynamical time and the spin period.
Model (A) exhibits begins with near-relativistic flow and only after $t\sim20$ s does the flow become
non-relativistic, with $\Gamma$ peaking at $\sim100$ at $t\sim6$ seconds.
We conclude that spindown in a Weber/Davis-type magnetic field geometry, taking $\eta=2$,
can be significant during the cooling epoch.  Even if a monopole-like field structure only obtains
during the $\Gamma<1$ non-relativistic epoch, an amount of energy comparable in magnitude with
the supernova explosion itself ($\gtrsim10^{51}$ erg) is extracted and emerges into the
post-explosion supernova ejecta.

{\bf Dipole Spindown:} We solve for $\Omega(t)$ in the same way as in the monopole case,
but here take $\eta=3$ in calculating $R_{\rm A}$ and, for $\Gamma>1$ use 
$\dot{E}_{\rm tot}=B_0^2R_\nu^6\Omega^4c^{-3}$.  
For a {\it slow} rotator with an initial period of 20 ms and $B_0=10^{15}$ G,
the final spin period is $\sim21$ ms.  That is, a $10^{15}$ G dipole field does not 
give significant spindown.  For this slow initial rotation period, only a 
dipole field greater than $10^{16}$ G affects the evolution significantly before $\Gamma>1$
and the simple vacuum magnetic dipole limit obtains.  As an extreme case, taking $B_0=5\times10^{16}$ G, the
final spin period is $\sim40$ ms, more than 70\% of the rotational energy is extracted
(a mere $\sim4\times10^{49}$ erg), and $\Gamma$ is greater than unity for $t>35$ seconds.
Figure \ref{plot:allps} shows the evolution of $P(t)$, $R_{\rm L}$, $R_{\rm A}$, $\Gamma$, 
and $E$ and $\dot{E}$. Virtually all of the braking occurs in these first 35 seconds.
As the wind transitions from non-relativistic to relativistic, spindown ceases on the timescale $\tau_{\rm KH}$
and the dipole limit obtains.

Figure \ref{plot:allpf} shows the time evolution computed for PNSs with initial spin period
of 1 ms and employing $\eta=3$.  For comparison, the same evolution is shown for three
different surface magnetic field strengths, $B_0=10^{16}$\,G, $5\times10^{15}$\,G, and $10^{15}$\,G,
labeled `A', `B', and `C', respectively.  Panels, linestyles, and units are the same as in Fig.~\ref{plot:allps}.
For $B_0=10^{15}$ G,  the final spin period is 1.15 ms.  Although the total change in spin period
is small,  approximately $10^{51}$ erg is extracted in the first second of the spindown calculation.  
A total of $\sim2\times10^{51}$ erg is ejected during the non-relativistic phase of
wind evolution ($t\lesssim9$ s), with $R_{\rm A}<R_{\rm L}$ and $\Gamma<1$.
Of course, for
the remaining 90 seconds of the spindown calculation, the simple vacuum dipole limit obtains.
For higher field strengths the transition from non-relativistic ($\Gamma<1$) to relativistic ($\Gamma>1$)
occurs at earlier times. For the highest field strength shown here (A), the magnetic energy density
dominates the kinetic energy density at $R_{\rm L}$ throughout the evolution.
Most important are our results from the (C) calculation, which show that 
even taking the more pessimistic dipole field ($\eta=3$ instead of $\eta=2$),
we find that an amount of energy comparable to the asymptotic supernova energy
is naturally injected into the post-shock supernova environment on a short timescale
with respect to the amount of time needed for the preceding supernova shockwave to traverse the progenitor
(see \S\ref{section:hypernovae}).

Note that by decreasing the initial spin period, while keeping $B_0=10^{15}$ G, the spindown
timescale drops sharply as a result of the exponential dependence of $\dot{M}$ on $\Omega$
(see eqs.~\ref{slowspin2} and \ref{mdotomega}).  
For example, taking an initial
period of 0.8 ms and $B_0=4\times10^{15}$ G, one finds $\dot{M}\simeq0.074$ M$_\odot$ s$^{-1}$
and $R_{\rm A}\sim17$ km. This implies a non-relativistic ($\Gamma<1$; see eq.~\ref{slowspinscale3})  
spindown timescale $\tau_J^{\rm NR}\sim2.6$ seconds.  
For comparison, the vacuum dipole spindown approximation (inappropriate
with such large $\dot{M}$)  yields $\tau_J^{\rm R}\sim31$ seconds.\footnote{Interestingly, if $B_0$ is increased, $R_{\rm A}$ increases, thus decreasing $\tau_J^{\rm NR}$.
However, for a given increase in $B_0$, $\tau_J^{\rm R}$ decreases more than $\tau_J^{\rm NR}$.
We find that $\tau_J^{\rm NR}/\tau_J^{\rm R}\propto B_0^{(4\eta-6)/(2\eta-1)}$.}
In the example given, with initial period 0.8 ms and $B_0=4\times10^{15}$ G only three seconds
transpire between the start of the calculation and the moment the vacuum dipole limit obtains.
However, in those few seconds, $\sim10^{52}$ erg is extracted -- a total of 0.019 M$_\odot$ 
with velocity $\sim1.5-3\times10$ cm s$^{-1}$.  Thus, even in the
more pessimistic dipole limit ($\eta=3$) a significant amount of rotational energy can be extracted
on a timescale much shorter than that inferred from a naive and, in this very early phase inappropriate,
application of vacuum dipole spindown.

\vspace*{.2cm}

\section{Discussion}
\label{section:implications}

\subsection{Hyper-Energetic Supernovae}
\label{section:hypernovae}

For the purposes of this paper we define the word `hypernova' as any supernova with an asymptotic
observed total energy of greater than $10^{51}$ erg.  We propose that the magnetically dominated,
centrifugally slung winds from rapidly rotating PNSs can explain hypernovae.

In this scenario, core-collapse is followed by shock stall and then explosion.
The explosion may be caused by a combination of neutrino heating and convection
(e.g.~Herant et al.~1994; Burrows, Hayes, \& Fryxell 1995; Fryer \& Warren 2002), 
by neutrino heating plus rotation and dissipative effects (Thompson, Quataert, \& Burrows in prep)
or perhaps via MHD-driven jets (Akiyama et al.~2003).  In any case,
the supernova shock propagates outward at $\sim10,000$ km s$^{-1}$ with energy $\sim10^{51}$ erg.  
The wind phase commences as the
PNS contracts and cools. In the first few seconds, the kinetic energy density and thermal pressure 
of the wind are large and the wind is not magnetically dominated.  
The wind hydrodynamical power in this phase is inefficiently supplied by neutrino heating as in NRNM wind models.
As the radius of the PNS decreases, the average magnetic field grows 
because of flux conservation ($B_0\propto R_\nu^{-2}$)
and an efficient dynamo as in Duncan \& Thompson (1992).  The PNS also spins up as a result of angular momentum
conservation. The hydrodynamical power of the
wind decreases with the neutrino luminosity and there is a point in time when the magnetic energy
density begins to dominate the thermal and kinetic wind energy density. 
An Alfv\'{e}n point will form, and efficient extraction of the 
rotational energy from the PNS will commence.  The mass loss rate will likely {\it increase} in this
phase despite the decrease in the neutrino luminosity as $R_{\rm A}$ approaches $R_{\rm sonic}$, assuming
that $R_\nu\Omega$ exceeds $c_s(R_\nu)$ (see \S\ref{section:nmcwind}).
This is a result of the exponential dependence of $\dot{M}$ on $\Omega^2$ in the centrifugal limit. 
For a rapidly rotating PNS, our calculations show that for $\eta$ close to 2,
of order $10^{52}$ erg is naturally extracted in the first few seconds.  
In \S\ref{section:extract} we showed that even in the case $\eta=3$, by appealing to slightly
higher $B_0$ and $\Omega$, of order $10^{52}$ erg can be extracted on a few-second timescale.
The velocity of this wind material evolves quickly to near $c$
and the NMC wind drives a secondary shock into the post-supernova-shock
material.  This secondary, but much more energetic shock encounters  the slower 
supernova shock while still inside the progenitor of a Type-II or Type-Ibc supernova
(probably at a radius $\lesssim10^5$ km).\footnote{Duncan \& Thompson (1992) showed
that magnetar spindown via vacuum dipole radiation can power hyper-energetic supernovae
in Type-II progenitors, but the spindown timescales via this mechanism are not
fast enough to explain hyper-energetic Type-Ibc supernovae.}  
In this way, a hyper-energetic supernova is created.

Only in a very compact progenitor will the secondary wind shock collide with the preceding
supernova shock outside the star.  This could plausibly occur in accretion-induced collapse 
of a white dwarf.  Otherwise, we expect this very energetic shock to alter the nucleosynthetic
signature of the hypernova with respect to normal supernovae, producing excess 
nickel as inferred from, e.g.~SN1998bw and SN2003dh 
(Nakamura et al.~2001a,b; Maeda et al.~2003;  Woosley \& Heger 2003).  If magnetars are born
in progenitors with extended envelopes, these strong magnetocentrifugal outflows might also 
prevent late-time fallback accretion (Chevalier 1989; Woosley \& Weaver 1995; Fryer et al.~1996).
Although the simple analysis presented here assumes sphericity, because the outflow is 
magnetocentrifugally driven, we expect asymmetries in the ejecta due to collimation
(e.g.~Sakurai 1985; Begelman \& Li 1994).

\subsection{Gamma Ray Bursts}

Although the time evolution is complex and depends sensitively on $B_0$, our models generally 
achieve large $\Gamma$ as $\dot{M}$ decreases at $R_{\rm L}$ (see middle lower panel in Fig.~\ref{plot:allp}).
Even though the velocity of the matter at $R_{\rm L}$ is only mildly relativistic, asymptotically the energy of the 
field may be transferred to the wind material, yielding ultra-relativistic velocities.
If all of the magnetic energy at $R_{\rm L}$
is eventually converted to kinetic energy, the asymptotic limiting Lorentz factor ($\gamma_\infty$) of the matter is 
$\Gamma$, given by eq.~(\ref{gammalim}).\footnote{Analyses of ideal relativistic winds predict that 
$\gamma_\infty\sim\Gamma^{1/3}$ (Michel 1969; Goldreich \& Julian 1970; Beskin et al.~1998).
However, modeling of the equatorial relativistic wind of the Crab pulsar (Kennel \& Coroniti 1984a,b;
Spitkovsky \& Arons 2003) suggests that $\gamma_\infty\sim\Gamma$ and {\it not}
that $\gamma_\infty\sim\Gamma^{1/3}$, in contradiction with theoretical expectations.}
Our exploration of the spindown of PNSs with 1 ms initial periods 
shows that $\Gamma$ may be very large with $\dot{E}$ in the range $\gtrsim10^{51}$ erg s$^{-1}$.  In fact, we 
obtain of order $\sim10^{51}-10^{52}$ erg emitted on a $10-100$ second timescale with $\Gamma\gtrsim100$
(Fig.~\ref{plot:allp}).  This conclusion is particularly robust for the optimistic 
monopole-like field geometry with $\eta=2$.  The correspondence between the 
 energy budget, energy loss rates, and Lorentz factors  inferred from observations
of GRBs and the models presented here support the conclusion that 
a class of PNSs, born highly magnetic and rapidly rotating within compact progenitors,
may be responsible for cosmological gamma ray bursts (Usov 1992; Thompson 1994). 

The now-robust association of long-duration GRBs with
highly energetic supernovae, as evidenced by SN1998bw (Galama et al.~1998) and 
SN2003dh (Stanek et al.~2003; Hjorth et al.~2003), follows naturally from our discussion of 
hypernovae in \S\ref{section:hypernovae}.  
As described in \S\ref{section:hypernovae},
we can expect complex shock structure evolution as the energy loss rate changes
and the outflow becomes increasingly more relativistic.  The dynamics of the supernova-shock/wind-shock 
interaction with the envelope and 
the associated instabilities might be important both in producing internal shocks and time
variability and ultra-high-energy cosmic rays (Arons 2003).  There are other ways to 
produce time variability.  We  find
that large changes in $\Gamma$ naturally occur as $R_{\rm L}$ evolves with $\Omega$ (see Fig.~\ref{plot:allp}).  
This order unity variability in $\Gamma$ provides a mechanism for producing internal shocks (Thompson 1994; 
Rees \& M\'{e}sz\'{a}ros 1994).  
Shocks in the flow can also be created by fluctuations in mass loading caused by sudden changes
in $\dot{M}$.  Closed loops that trap wind material
on the surface of the PNS may be sheared open on millisecond timescales due to convection
or differential rotation or may be opened by continued neutrino heating of the trapped matter (Thompson 2003).
Finally, relativistic MHD instabilities (e.g.~reconnection) may produce time-variability
in Poynting-flux dominated outflows.  

The duration of gamma ray bursts in the magnetar model presented here
is somewhat uncertain and could be set by either the spindown time of
the PNS ($\tau_J$) or the PNS cooling time $\tau_{\rm KH}$.\footnote{In
principle the duration could also be set by much more 'messy' physics
such as the detailed emission model and how it depends on $\dot{E}$
and $\Gamma$ in the wind.}  The latter would be appropriate if $\tau_J
> \tau_{\rm KH}$ and if the magnetic field significantly decreases
after the cooling epoch, leading to a decrease in wind luminosity and
thus an end to a detectable GRB.  The field at the Alfven point
(relevant for the spindown power) could decrease either because the
{\it surface} field decreases in strength as convection in the
PNS ceases (Thompson 1994)  or because the field may transition
from monopole-like to dipole-like as $\dot{M}$ decreases.

A further important feature of the magnetar model presented here is that in the monopole case,
$\dot{E}$ decreases {\it exponentially} as a function of $t$ with timescale given by eq.~(\ref{relspinscale2}).
This is a prediction of the monopole-like magnetar model and it stands
in sharp contrast to the energy loss rate predicted from both collapsar-type models (Woosley 1993)
whose $\dot{E}$ must evolve on the timescale for accretion onto the central compact object.
Equation (\ref{relspinscale2}) also shows that if the field is increased to $10^{16}$ G that $\tau_{\rm J}^{\rm R}\sim0.3$ 
seconds, which is in the middle of the distribution for short-duration GRBs.  
Similarly, allowing for lower $B_0$ and $\Omega$, we further speculate that if magnetars are responsible for long-duration GRBs, 
that much longer ($\gtrsim1000$ s) GRBs may
exist and could be detected by upcoming observations with improved sensitivity (e.g.~SWIFT).
If one allows for a dynamo mechanism operating efficiently within the
convective core of protoneutron stars (Duncan \& Thompson 1992; Thompson \& Duncan 1993)
then $B_0$ is directly proportional to $\Omega$ to some power, depending on the details of the dynamo.
If there is such a relationship between $B_0$ and $\Omega$, then the diversity of GRB sources, 
both long and short GRBs (Kouveliotou et al.~1993) as well as hyper-energetic
supernovae, may be grouped in a one-parameter family, a function only of rotation period.
In this way, the birth of rapidly rotating PNSs may be responsible for a diverse variety of astrophysical explosions.
We speculate that short bursts come from more rapidly rotating progenitors with no extended envelope.
For example, the short GRB population may
be due to accretion-induced collapse of a white dwarf (leading to a sub-millisecond PNS), 
whereas the long population is from collapse of Type-Ibc progenitors (perhaps only $\sim1$ ms periods).  
Another possibility for short GRBs is that the merger of a neutron star binary creates a black hole or massive
rapidly rotating neutron star that drives a wind similar to those detailed in this paper (Thompson 1994).
The scaling relations derived here are applicable in this context as well for magnetocentrifugally
driven flows off the accretion torus produced from the inner regions of a black hole accretion disk.  
In either case, the short bursts would trace an old stellar
population, whereas the long bursts would be associated with star-forming regions.

That the birth of neutron stars with magnetar-like field strengths might be the astrophysical
origin of GRBs is not new.  The idea presented here follows that of Usov (1992) and Thompson (1994)
and touches on any model of GRBs requiring Poynting-flux dominated outflows (e.g.~Lyutikov \& Blandford 2003).
In Usov (1992) and Thompson (1994), only a pure dipole field geometry was considered.  In this paper we find potentially
much larger energy loss rates for a given $B_0$ and $\Omega$ as a consequence 
of different field geometries.  In particular, motivated by the literature on stellar winds and
their angular momentum evolution, we have considered monopole field topologies in the spirit of Weber \& Davis (1967).
Further consideration of rapid rotation coupled with high mass loss rates (see \S\ref{section:eta};
Mestel \& Spruit 1987) supports this view.  The energy extraction rates and braking
suggested by our pure monopole calculations (see \S\ref{section:analytic} and \S\ref{section:extract})
 are likely overly optimistic. However, we await multi-dimensional MHD models of relativistic outflows.

In addition to the issue of field topology there are other important potential difficulties inherent to this
GRB mechanism.  Similar to the collapsar model of GRBs (Woosley 1993; Macfadyen \& Woosley 1999),
the energy injection rate must remain high on the timescale of the relativistic ejecta to break out of the progenitor. 
Given the timescales we have derived and $\tau_{\rm KH}$, this is plausible for compact Type-Ibc core-collapse 
scenarios.  In addition, we rely on conversion of magnetic
energy into kinetic energy asymptotically, a process that is not well understood in models 
of pulsar winds.  Issues of collimation and -- possibly -- jet formation require multi-dimensional
models and we assume that many of the structures generic to collapsar jets will
obtain  here as well (Aloy et al.~2000; Zhang et al.~2003), although the flow will be Poynting-flux dominated.  

\subsection{Conclusions}

We have estimated spindown timescales for protoneutron stars born rapidly rotating
and with large surface magnetic field strengths.  The combination of magnetocentrifugal effects and
neutrino energy deposition combine to produce a wind capable of significantly
effecting the early angular momentum evolution of newly born protoneutron stars.
In addition, we have constructed steady-state models
of neutrino-magnetocentrifugal winds,  including magnetic fields implicitly and approximately 
by forcing the wind matter into strict corotation with the PNS.  

\begin{figure}
\begin{center}
\includegraphics[height=15cm,width=15cm]{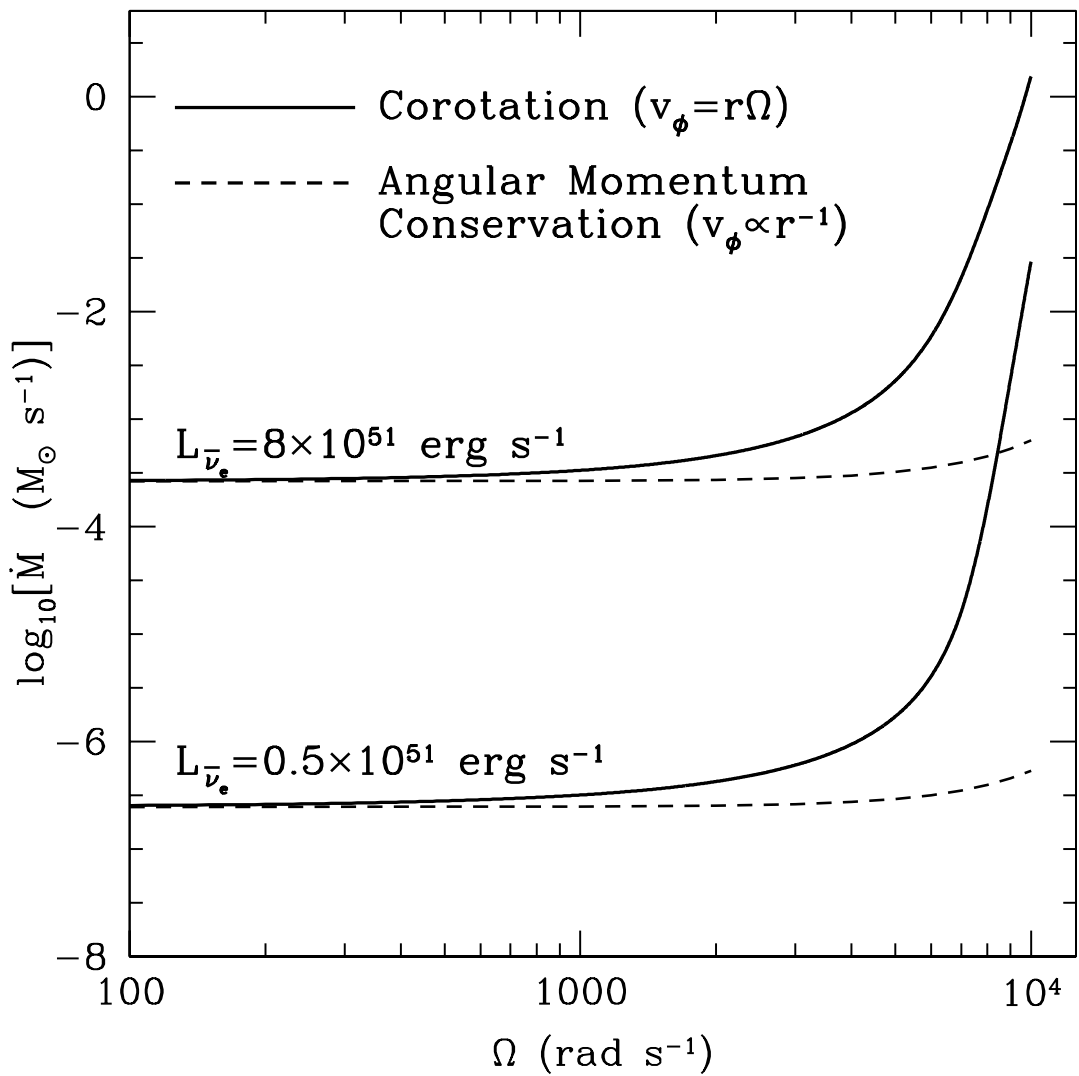}
\end{center}
\caption{Log of the mass loss rate as a function of $\Omega$ for 
$L_{\bar{\nu}_e}=8\times10^{51}$ erg s$^{-1}$ 
and $L_{\bar{\nu}_e}=0.5\times10^{51}$ erg s$^{-1}$ assuming corotation with the
PNS surface, $v_\phi=r\Omega$ (solid lines).  Also shown are results assuming
no corotation (dashed lines), but taking $v_\phi=R_\nu\Omega(R_\nu/r)$, appropriate
for angular momentum conservation.  For $R_\nu\Omega>c_s(R_\nu)$, $\dot{M}$ increases
approximately exponentially with $\Omega^2$ in the models assuming corotation.
}
\label{plot:mdotthis}
\end{figure}

\begin{figure}
\begin{center}
\includegraphics[height=15cm,width=15cm]{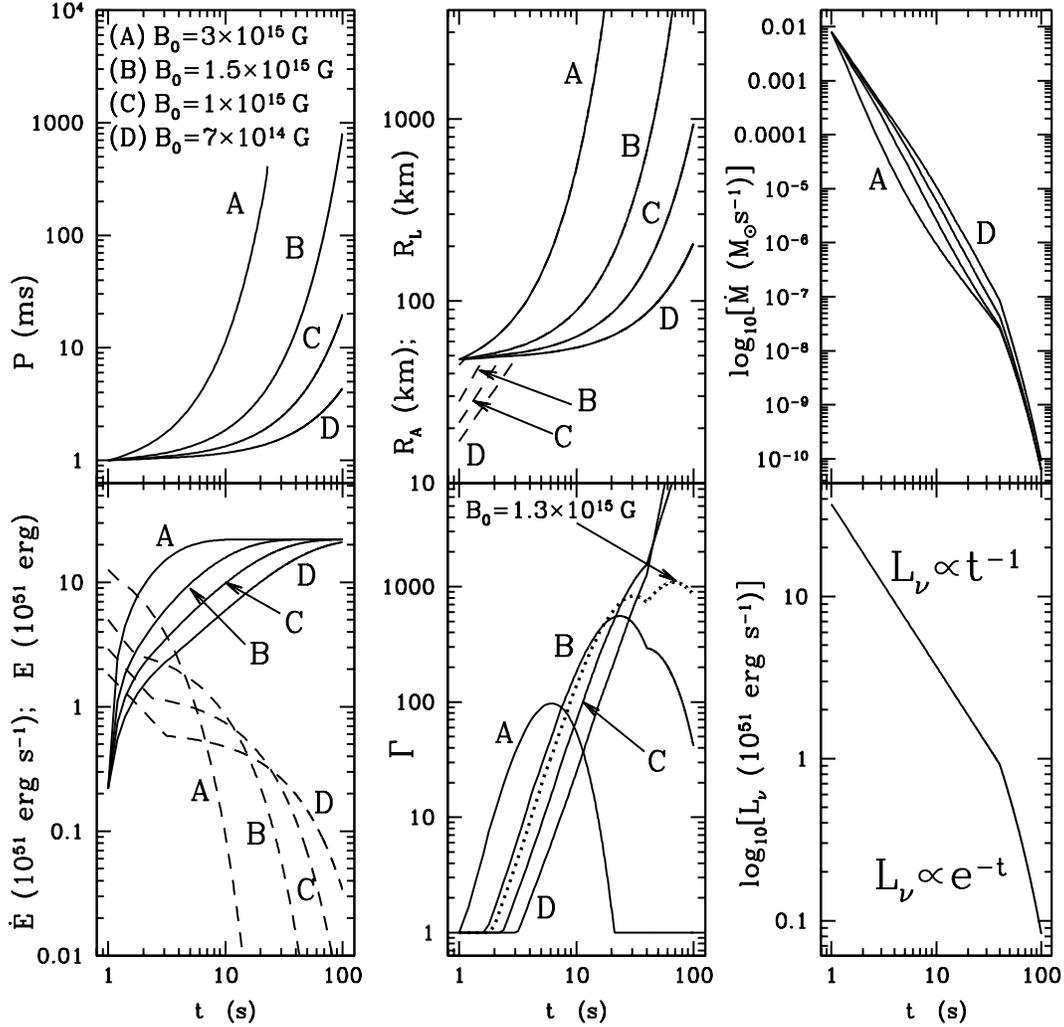}
\end{center}
\caption{Representative time evolution for the spin evolution of a rapidly rotating highly magnetic PNS
with a monopole field geometry ($\eta=2$).  Calculations with four different surface magnetic field strengths are shown: 
$B_0=3\times10^{15}$ G (A), $1.5\times10^{15}$ G (B), $10^{15}$ G (C), and $7\times10^{14}$ G (D).  
The upper and lower right panels
show the mass loss rate ($\dot{M}$ [M$_\odot$ s$^{-1}$]) and the total neutrino luminosity 
($\log_{10} L_\nu$ [$10^{51}$ erg s$^{-1}$]), respectively.  The spin period
($P$ [ms]; upper left panel), limiting asymptotic Lorentz factor ($\Gamma$; lower middle panel),
spin energy extracted ($E$ [$10^{51}$ erg]; lower left panel, solid lines), and 
energy extraction rate ($\dot{E}$ [$10^{51}$ erg s$^{-1}$]; lower left panel, dashed lines), are shown.
In addition, the upper-middle panel shows $R_{\rm L}$ (solid lines) and $R_{\rm A}$
as computed with $\eta=2$ (dashed lines).   For comparison, in the plot of $\Gamma$ versus $t$,
we also include the evolution for $B_0=1.3\times10^{15}$ G (dotted line), which falls between
models (B) and (C).  Note the interesting time evolution of $\Gamma$ as  $\dot{M}$ drops exponentially
after $t=40$ s, in response to the assumed change in $L_\nu$.}
\label{plot:allp}
\end{figure}

\begin{figure}
\begin{center}
\includegraphics[height=15cm,width=15cm]{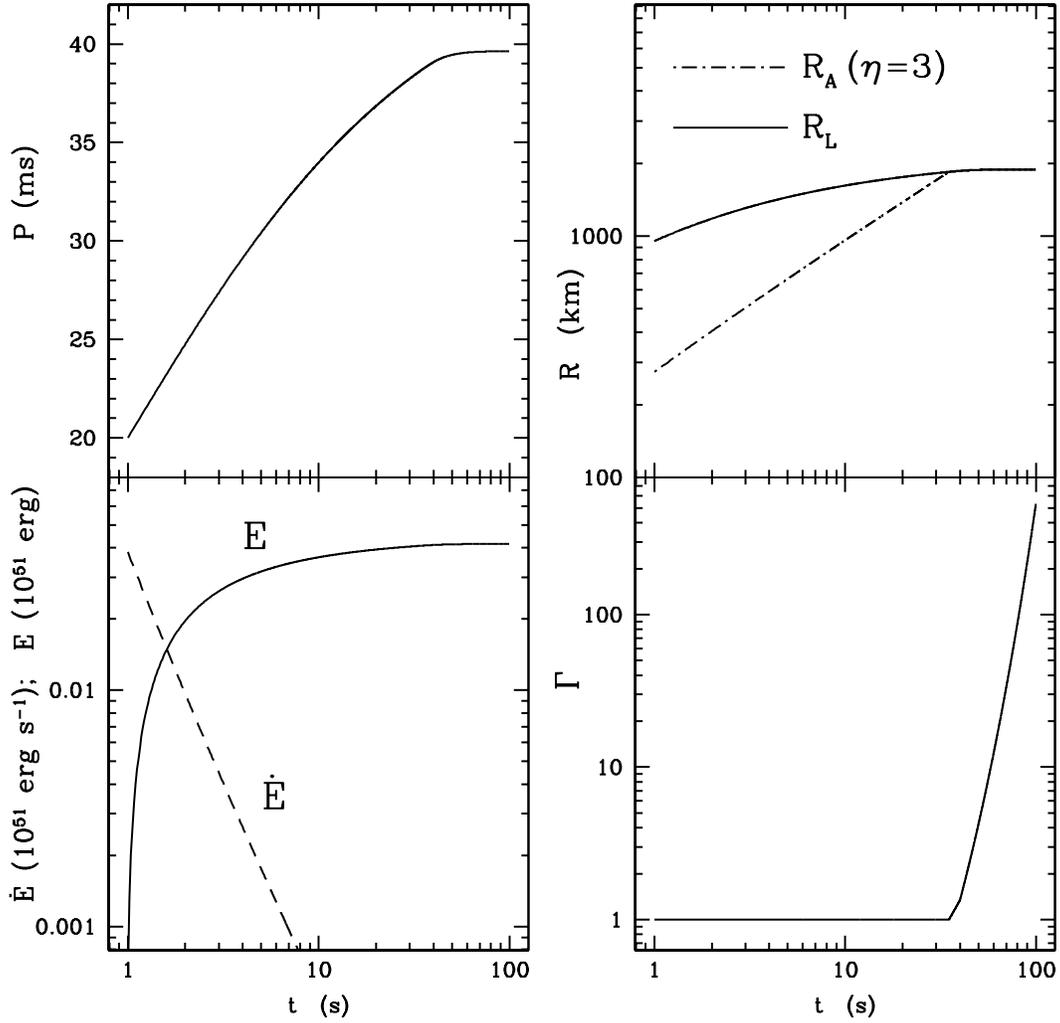}
\end{center}
\caption{Representative time evolution for the spin evolution of a slowly rotating highly magnetic PNS
with a dipole field geometry ($\eta=3$).  
Spin period
($P$ [ms]; upper left panel), limiting asymptotic Lorentz factor ($\Gamma$; lower right panel),
light cylinder radius ($R_{\rm L}$ [km], upper right panel, solid line),
Alfv\'{e}n radius ($R_{\rm A}$ [km], upper right panel, dot-dashed line),
spin energy extracted ($E$ [$10^{51}$ erg]; lower left panel, solid line), and
energy extraction rate ($\dot{E}$ [$10^{51}$ erg s$^{-1}$]; lower left panel, dashed line), are shown.
As an extreme example, $B_0$ was taken to be $5\times10^{16}$ G.}
\label{plot:allps}
\end{figure}

\begin{figure}
\begin{center}
\includegraphics[height=15cm,width=15cm]{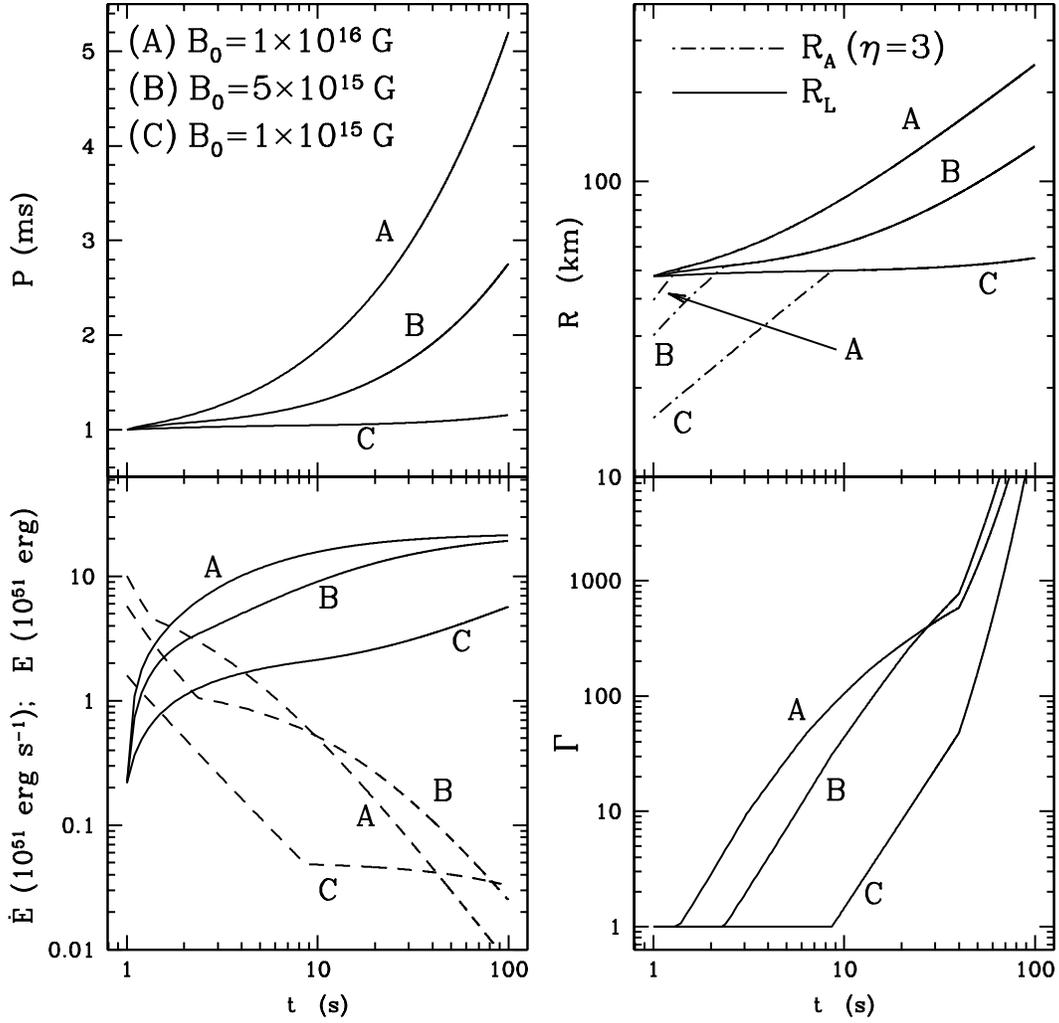}
\end{center}
\caption{Representative time evolution for the spin evolution of a rapidly rotating highly magnetic PNS
with a dipole field geometry ($\eta=3$).  Calculations with three different surface magnetic field strengths are shown: 
$B_0=10^{16}$ G (A), $5\times10^{15}$ G (B), and $10^{15}$ G (C).  Panels, units, and linestyles
are the same as in Fig.~\ref{plot:allps}.}
\label{plot:allpf}
\end{figure}

We find that magnetar-like surface fields ($\sim10^{15}$ G) dominate the thermal pressure
and kinetic energy density of PNS winds in the first few seconds after the preceding 
supernova for a wide range of initial spin periods.  We show that as the mass loss from the
PNS abates, the outflow becomes increasingly faster, transitioning from non-relativistic
to relativistic outflow  in the first few to tens of seconds.  We explore this progression and 
provide evolutionary models that bridge the transition.  We find that non-relativistic
spindown timescales can be considerably shorter than those inferred from simple
vacuum dipole spindown.  This results both from physics of the Alfv\'{e}n point and
the large enhancement in mass loss rate expected in models with spin periods less than $\sim2$ ms
(see \S\ref{section:nmcwind} and Fig.~\ref{plot:mdotthis}).  Motivated by the literature
on stellar winds from magnetic rotating stars, we have carried out calculations for
monopole-like field geometries in addition to pure dipole fields (see \S\ref{section:eta} and \S\ref{section:nmcwind}).
Although optimistic in computing spindown, quasi-monopolar fields are physically motivated by 
the large mass loss and rapid rotation of the systems considered.  We are currently working on
axi-symmetric two-dimensional MHD models to more fully explore the issue of global magnetic field topology
and angular momentum loss.

Regardless of field topology, we find generically in our models that if magnetars are born with $\sim1$ ms
initial periods and $B_0\sim10^{15}$ G that more than $10^{51}$ erg of rotational energy can be extracted 
in the first few seconds and that the velocity of this matter will approach $c$ as the Alfv\'{e}n radius approaches the
light cylinder.  As the mass loss rate continues to plummet, the asymptotic limiting Lorentz factor of the
ejected matter may approach several hundred.  We therefore propose these objects as candidate central 
engines for hyper-energetic supernovae and cosmological gamma ray bursts of both the long and,
possibly, the short variety (see also Usov 1992, Thompson 1994, and
Lyutikov \& Blandford 2003).  

Although more mundane from an energetic point of view, if monopolar field geometries 
are at all relevant for slow rotators ($P\gtrsim20$ ms), efficient angular momentum loss results.
We find that if magnetars are born with the spin periods of normal neutron stars, taking $\eta=2$,
that a PNS can be spun down from a spin period of $\sim50$ ms to a spin period of $\sim1$ second
in $\sim10$ seconds of wind evolution.

We emphasize that for any field strength there is a point in time during the cooling of the
nascent neutron star when the magnetic field dominates the wind dynamics.
If the field is strong enough, this happens early in the wind phase and efficient extraction
of rotational energy is likely a natural consequence. The neutrinos still carry away the
neutron star's binding energy ($\sim3\times10^{53}$ erg), but their energetic coupling with
the surrounding environment is (necessarily) weak.  In marked contrast, the rotational
energy (although only of order $10^{52}$ erg for a fast rotator) may be efficiently coupled
to the wind material via the strong magnetic field.  It is in this way that the wind's 
energetic presence within supernova remnants and in the spin distribution of neutron stars
may be observed.

\acknowledgments

We thank Lars Bildsten for originally suggesting this problem and for helpful conversations on 
potential observational implications.  
We gratefully acknowledge conversations with Jonathan Arons, Anatoly Spitkovsky, Yoram Lithwick,
Brian Metzger, and Adam Burrows.  We are also indebted to Chris Thompson and Henk Spruit for a critical reading of the text
and for helpful comments and suggestions for improvement.
We thank Evonne Marietta for making her general electron-positron EOS available.
T.A.T. thanks the Aspen Center for Physics, where much of this work germinated.
T.A.T. is supported by NASA through Hubble Fellowship
grant \#HST-HF-01157.01-A awarded by the Space Telescope Science
Institute, which is operated by the Association of Universities for 
Research in Astronomy, Inc., for NASA, under contract NAS 5-26555.
E.Q. is supported in part by NSF grant AST 0206006, NASA grant NAG5-12043,
an Alfred P. Sloan Fellowship, the David and Lucile Packard Foundation, and a Hellman Faculty Fund Award.

\clearpage

\clearpage

\clearpage

\clearpage

\clearpage

\end{document}